\documentclass[preprint, 12pt, number, square, comma, sort&compress]{elsarticle}

\usepackage{amssymb}
\usepackage{color}
\usepackage{multirow}
\usepackage[utf8]{inputenc}
\usepackage{verbatim}
\usepackage{hyperref}
\usepackage{subfigure}
\usepackage{caption}
\usepackage{booktabs}
\usepackage{gensymb}
\usepackage{float}
\usepackage{amsmath,bm}
\usepackage[version=4]{mhchem}
\usepackage{siunitx}
\usepackage{longtable,tabularx}
\journal{Aerospace Science and Technology}

\begin{document}

\begin{frontmatter}



	\title{An implicit coupling framework for numerical simulations between hypersonic nonequilibrium flows and charring material thermal response in the presence of ablation}


	\author[1,2]{Jingchao Zhang}
	\author[3]{Chunsheng Nie}
	%\ead{zhangjingchao@mail.nwpu.edu.cn}
	\cortext[cor1]{Corresponding author}
	\author[1,2]{Jinsheng Cai}
		% \ead{caijsh@nwpu.edu.cn}
	\author[1,2]{Shucheng Pan\corref{cor1}}
	\ead{shucheng.pan@nwpu.edu.cn}
	\address[1]{School of Aeronautics, Northwestern Polytechnical University, Xi'an 710072, China}
	\address[2]{National Key Laboratory of Aircraft Configuration Design, Xi'an 710072, China}
	\address[3]{Science and Technology on Space Physics Laboratory, China Academy of Launch Vehicle Technology, Beijing 100076, China}

	%\affiliation{organization={NWPU},%Department and Organization

	\begin{abstract}
		An implicit coupling framework between hypersonic nonequilibrium flows and material thermal response is proposed for the numerical simulation of ablative thermal protection materials during its flight trajectory. Charring ablative materials, when subjected to aerodynamic heating from hypersonic flows, undergo complex processes such as ablation and pyrolysis, involving heterogeneous and homogeneous chemical reactions. These multi-physical phenomena are simulated by a multicomponent material thermal response (MTR) solver that takes into account the complexity of component of pyrolysis gases. The species concentrations are calculated to improve the accuracy of transport and thermophysical parameters of pyrolysis gases. The MTR solver implements implicit time integration on finite difference discretization form to achieve higher efficiency. The numerical solutions of hypersonic flows and material thermal response are coupled through a gas-surface interaction interface based on surface mass and energy balance on the ablating surface. The coupled simulation employs the dual time-step technique, which introduces pseudo time step to improve temporal accuracy. The explicit coupling mechanism updates the interfacial quantities at physical time steps, which achieves higher computational efficiency, but introduces time discretization errors and numerical oscillations of interfacial quantities. In contrast, the implicit coupling mechanism updates the interfacial quantities at pseudo time steps, which reduces the temporal discretization error and suppresses numerical oscillations, but is less efficient. In addition, a simplified ablation boundary based on steady-state ablation assumption or radiation-equilibrium assumption is proposed to approximate solid heat conduction without coupling the MTR solution, providing quasi-steady flow solutions in the presence of ablation. 
	\end{abstract}


	%Research highlights
	\begin{highlights}
		\item A multicomponent material thermal response solver for numerical simulation methods of charring ablators is developed.
		\item An ablation boundary is proposed to solve quasi-steady flows in the presence of surface ablation.
		\item Numerical solutions of hypersonic flows and material thermal response are implicitly coupled by exchanging interfacial quantities at pseudo time step to achieve higher accuracy.
		% \item The coupled simulation build a gas-surface interaction model based on surface mass balance and surface energy balance to account for mass injection and energy dissipation on the ablating surface.
		\item Implicit coupling mechanism suppresses numerical oscillations of interfacial quantities and achieves higher time discretization accuracy.
	\end{highlights}


\end{frontmatter}

\section{Introduction}% and search for habitable environment防热材料的热设计是非定常过程，目前大部分方法实现流场与MR的解耦计算或近似模型逼近，而没有完全模拟热响应与流场的双向耦合机制during the entry, descent and landing (EDL) sequence With an ever-increasing interest in understanding the nature of our solar system, humanity has embarked on a major endeavor in space exploration.
Space exploration missions face severe aerodynamic heating challenges during the entry, descent, and landing (EDL) sequence. This heating arises from the conversion of kinetic energy in hypersonic flows due to viscous dissipation and shock compression, causing chemical reactions and excitation of vibrational energy within high temperature gas mixtures~\cite{anderson2000hypersonic}. To mitigate the heat load, spacecraft requires thermal protection system (TPS) on the surface to ensure safe operation temperature of internal modules. TPS materials can be broadly categorized into reusable and ablative type. Reusable materials, such as Alumina Enhanced Thermal Barrier~\cite{10.1002/9780470320402}, usually have high thermal resistance and remains mechanically and chemically stable during flight mission. In contrast, ablative materials provides enhanced thermal protection by carrying away high enthalpy gases through ablation and pyrolysis. Ablation is surface reactions involving heterogeneous reactions and phase changes, resulting in surface recession. Pyrolysis is the endothermic decomposition of resin matrix within charring ablative materials, exhausting high-enthalpy gases to release heat flux. Ablative materials can be classified as charring and non-charring ablators based on whether pyrolysis occurs or not. Charring materials consist of carbon fibers impregnated with phenolic resin matrix, which includes pore structures that provides higher mass efficiency for thermal protection. 
%这段补上ablative material的数值方法 This method maintains the grid cell unchanged while introducing a source term due to the moving system, thus is easy to extend to multidimensional problems. Amar et al.~\cite{doi:10.2514/1.29610} developed the CVFEM to achieve second-order spatial accuracy and first-order temporal accuracy with a systematic verification process. 

Numerical methods for modeling material thermal response (MTR) of ablative materials have been developed rapidly. The first numerical model of ablative materials was proposed by Moyer and Rindal~\cite{Moyer1968} in the 1960s through a one-dimensional finite difference Charring Material Thermal response and Ablation program (CMA). In 1988, Blackwell et al.~\cite{doi:10.1080/10407788808913631} introduced the control volume finite element method, which iteratively solves the energy equation in a moving control volume attached to the receding surface. In 1990, NASA Ames Research Center proposed several improvements to enhance numerical stability and developed the Fully Implicit Ablation and Thermal Response Program (FIAT)~\cite{doi:10.2514/2.3469}. However, these numerical methods introduce serval simplifications that may compromise accuracy, such as neglecting the momentum equation of pyrolysis gases within porous materials. The present work builds a comprehensive numerical framework for numerical simulations of ablative materials, incorporating recent advances in theoretical modeling~\cite{annurev-fluid-030322-010557}. It introduces species conservation and implicit time integration to improve accuracy and efficiency~\cite{LACHAUD20151034}. Scandelli~\cite{SCANDELLI2023108297} proposed a two temperature material response model based on thermal and chemistry non-equilibrium to solve the gas phase and solid phase temperature in charring materials.%CMA addressed the energy balance equation, albeit with several simplifying assumptions, and employed a node-dropping procedure to move mesh points.A Newton's method including a sensitivity matrix is implemented to achieve second-order nonlinear convergence. 

 %其一是流场求解在边界上使用热化学平衡边界（B table）来近似壁面反应和能量，其二是用边界层方程近似壁面换热量
In order to simulate the thermal response of ablative materials during flight missions, it is necessary to couple the numerical simulation methods of hypersonic flows and ablative materials. Generally, numerical methods of hypersonic flows solve compressible multicomponent reactive Navier-Stokes equations. The hypersonic flows provide heat flux boundary for the MTR solver. While, the MTR solver solves the wall temperature variant and pyrolysis gas velocity, which are the ablative wall boundary conditions of hypersonic flows. Historically, the coupled simulation employs uncoupled or loosely coupled methods, which ignore or approximate the coupling mechanism. The widely used assumption is the thermochemical balance model (B' table)~\cite{An1968}, which assumes chemical equilibrium on the ablating surface and computes the dimensionless ablation mass flux through the minimization of Gibbs free energy. B' table typically incorporates with simplified boundary layer equations to approximate the mass and energy transfer on the ablating surface, avoiding the solution of time-consuming Navier-Stokes equations~\cite{lachaud2015detailed}. Another kind of loosely coupled approach~\cite{BIANCHI2021121430,doi:10.2514-1.47995} solves quasi-steady ablative flows by approximating the MTR solutions with a quasi-steady state assumption~\cite{10.1115/1.4034595} or a radiative equilibrium assumption~\cite{bianchi2007numerical}. Turchi et al.~\cite{TURCHI201325} investigated uncertain parameters in the numerical simulation of carbon–phenolic solid rocket motor nozzle through steady-state ablation assumption. These assumptions contain serval simplifications that can compromise accuracy. Hence, strong coupling approaches~\cite{johnston2012study} including detailed surface mass balance and surface energy balance on the ablating surface have been developed to achieve higher accuracy. Başkaya et al.~\cite{BASKAYA2024106134} evaluated the accuracy of immersed boundary methods to model surface ablation, including the influence of chemical nonequilibrium and gas–surface interactions.%This method assumes a constant Lewis number and introduce a correction coefficient to account for the blockage effect~\cite{reynier2013survey}

 %第二部分是迭代误差，讲当前的计算方法和不足，难点（时间尺度问题，耦合误差），引出本文的目的和解救方法
Most strong coupling methods traditionally use a explicit coupling mechanism by exchanging and updating the interfacial properties at specific physical time level while freezing the ablation boundary condition. Conti et al.~\cite{conti1992practical} solves Navier-Stokes equations coupled with material response model to yield transient solutions for carbon-based TPS. Appar et al.~\cite{doi:10.1063-5.0082783} conducted a conjugate thermal analysis for charring ablative materials through coupling a direct simulation Monte Carlo flow solver with a material thermal response solver. Martin et al.~\cite{doi:10.2514/1.A32847} proposed a strongly coupled blowing boundary in arbitrary Lagrangian-Eulerian (ALE) system to describe the recession caused by surface ablation. However, these explicit coupling mechanisms introduce numerical instabilities due to the abrupt change of interfacial quantities and time discretization errors. To address this issue, Kuntz et al.~\cite{doi:10.2514/2.6594} proposed an estimation-correction iteration to reduce the instabilities by interpolating surface quantities during the coupling procedure. Chen et al.~\cite{CHEN2018620} proposed an adaptive time-step approach using a proportional-integral-derivative (PID) controller to select the time-step adaptively, which improves efficiency for the simulation of coupled fluid-thermal-structural behaviors of hypersonic wings. Zibitsker et al.~\cite{ZIBITSKER2024124728} developed a coupled framework between hypersonic flows and material response with overset grid, showing good agreement with experimental data. %This approach was further improved by Cross et al.~\cite{doi:10.2514/1.T5860}, who introduced an under-relaxation factor to blend the interfacial quantities between previous and current values.

This manuscript proposes an implicit coupling framework that simultaneously solves the hypersonic flows and material thermal response to reduce the time discretization errors and suppress the oscillations of wall quantities. The manuscript is motivated by three main objectives: the first is to develop robust numerical methods for modeling material thermal response in the presence of ablation and pyrolysis, the second is to propose a gas-surface interaction interface with detailed modeling of mass and heat transfer on the ablating surface, and the last is to improve the accuracy and efficiency of the numerical methods in coupled simulations of ablative materials subjected to aerodynamic heating during the flight trajectory. %encounter considerable challenges due to the disparity in characteristic time scales between turbulence, heat transfer, and chemical reactions. These discrepancies result in numerical stiffness and convergence issues

\section{Theoretical and numerical framework}
\subsection{Hypersonic nonequilibrium flows}

The governing equations of hypersonic flows are compressible multicomponent reactive Navier-Stokes equations. The two temperature model~\cite{park1989assessment} is employed to account for thermodynamic nonequilibrium, using a translation-rotation temperature $T$ describing the heavy-particle translational and rotational energy, and a vibration temperature $T_v$ describing the vibrational energy. The conservation equations of mass, momentum, and energy are written as,%, the electronic, and the electron-particle translational

\begin{equation}
	\label{Eq:govern}
	\frac{{\partial {\bm{U}}}}{{\partial t}} + \nabla  \cdot \left( {{\bm{F}} - {{\bm{F}}_d}} \right) = {\bm{S}},
\end{equation}
where $\bm{U} = [\rho{\bm{Y}^{\rm{T}}},\rho{\bm{u}^{\rm{T}}},\rho e,\rho e_v]^{\rm{T}}$ and $\bm{S} = [\bm{\dot \omega}^{\rm{T}}, \bm{0}^{\rm{T}},0,\dot \omega_v]^{\rm{T}}$ are the vector of conservative variables and source term, respectively. $\nabla = [\partial_x,\partial_y,\partial_z]$ is the Hamiltonian operator. ${\rho}$ is the density, ${\bm{Y}} = [{Y_1}, \ldots ,{Y_{ns}}]$ is the vector of the mass fraction of the individual species, and ${\bm{u}} = [u,v,w]$ is the bulk velocity components. $e$ and $e_v$ are the specific total energy and the specific vibrational energy, respectively. $\bm{\dot \omega} = [{\dot \omega}_1,\ldots,{\dot \omega}_{ns}]$ is the vector of the net production rate of the individual species, and $\dot\omega_v$ is the vibrational energy source evaluated by the Landau-Teller model~\cite{MillikanSys}.

The convective fluxes and viscous fluxes are expressed by
\begin{equation}
	\label{Eq.flux}
	{\bm{F}} = \left[ \begin{array}{l}
		\rho {{\bm{Y}}^T}                    \\
		\rho {{\bm{u}}^{\rm{T}}}\\
		(\rho e + p)                              \\
		(\rho e_{v})
	\end{array} \right]{\bm{u}} + 
	\left[ \begin{array}{l}
		\bm{0}                    \\
		{\bf{I}}p \\
		\bm{0}                              \\
		\bm{0}
	\end{array} \right]
	,
	{{\bm{F}}_d} = \left[ \begin{array}{l}
		- {\bm{J}}                                                                    \\
		{\bm{\tau }}                                                                  \\
		{\bm{u\tau }} - {{\bm{q}}} - {{\bm{q}}_{v}} - ({{\bm{h}}^{\rm{T}}}{\bm{J}}) \\
		- {{\bm{q}}_{v}} - ({\bm{e}}_{v}^{\rm{T}}{\bm{J}})
	\end{array} \right],
\end{equation}
respectively, where ${p}$ is the pressure, ${\bm{h}}$ is the vector of species specific enthapy, and ${{\bm{e}}_{v}}$ is the vector of species specific vibrational energy. ${\bm{q}}$ and ${\bm{q}_{v}}$ are the directional heat flux vector, i.e., $\bm{q} = -\lambda \nabla T$ and ${\bm{q}_{v}} = {-\lambda _v}\nabla T_v$, where ${\lambda}$ and ${\lambda _v}$ are the thermal conductivity of $T$ and $T_v$, respectively. $\bm{J}$ is the directional species' diffusion flux tensor computed by Fickian diffusion law, i.e., $\bm{J_{s}} = \rho {D_s}\nabla Y_s$, where ${D_s}$ is the effective diffusion coefficient of the $k$-th specie. The thermodynamic and transport properties of mix gases are computed by the Wilke's mixing rule with the pure species obtained by the NASA-9 curve fits~\cite{gupta1990review} and Blottner's curve fits~\cite{blottner1971chemically}, respectively. %The shear stress tensor is computed by, $\bm{\tau} = 2\mu (\bm{S} - \bf{I}(\nabla \cdot \bm{u})/3 ) + \mu'\bf{I}\nabla \cdot \bm{u}$, where $\bm{S}$, $\mu$ and $\mu'$ are the strain rate tensor, the first and second dynamic viscosity, respectively. 

% More detailed derivation of chemical reaction source, vibrational excitation, and thermodynamic properties can refer to our previous work~\cite{ZHANG202489,10.1063/5.0218188}, which also provides validation cases for the numerical methods of hypersonic non-equilibrium flows.

The governing equations are solved by our finite-volume in-house chemically reacting flow (CRF) solver, the accuracy of which has been validated in our previous work~\cite{ZHANG202489,10.1063/5.0218188}.% The MUSCL scheme~\cite{VANLEER1979101} with the minmod limiter~\cite{SwebyHigh} is applied to achieve a shock capturing capability. %The 11 species ($\rm{N_2}$, $\rm{O_2}$, $\rm{NO}$, $\rm{C_3}$, $\rm{CO_2}$, $\rm{C_2}$, $\rm{CO}$, $\rm{CN}$, $\rm{N}$, $\rm{O}$, $\rm{C}$) with 18 reactions~\cite{doi:10.2514/1.J052659} is used for pure ablation employs  is taken from Ref.he convective flux is splitted by the AUSMPW+ scheme~\cite{kim2001methods}The chemical reaction mechanism contains 24 reactions between the ablation products and air~\cite{doi:10.2514/1.J052659}. 

\subsection{Charring material thermal response}
When the temperature of the ablative material increases, the polymer matrix undergoes decomposition, with the pyrolysis gas being exhausted to carry away heats, leaving behind carbon fibers. Based on the volume fraction of polymer matrix, $c=((\rho - \rho_c)/((\rho_v - \rho_c)))$, the charring material is divided into three layers: the char layer $(c<0.02)$, the pyrolysis layer $(0.02<c<0.98)$, and the virgin layer $(c>0.98)$. The governing equations for material thermal response (MTR) of charring ablative materials are written as multicomponent volume-averaged multiphase conservation equations of mass, momentum, and energy.
\subsubsection{Mass conservation equation}
The mass conservation equation of the $s$-th specie is written as
\begin{equation}
	\label{Eq.mass}
	\frac{\partial( \epsilon \rho_s)_{pg}}{\partial t}  + \nabla \cdot ((\bm{u} \epsilon \rho_s)_{pg}) = \dot{m}_{pg,s},
\end{equation} 
where $\epsilon$ is the volume fraction of the pyrolysis gas ($pg$), $\rho_s$ and $\dot{m}_{pg,s}$ are the partial density and the production rate of the $s$-th species, respectively. $\dot{m}_{pg,s}$ is traditionally obtained by fitting thermogravimetric analysis. The decomposition of the fictitious sub-phase $j$ from the solid phase $i$ of polymer matrix ($pm$) yields $Z_s$, i.e.,
\begin{equation} 
	pm_{i,j} \rightarrow \sum_{s\in[1,ns]}\theta_{i,j,s}Z_s,\quad \forall i \in [1,N_p],\forall j \in [1,P_i]
\end{equation}
where $N_p, P_i$ is the total number of the solid phase and the sub-phase, respectively. The production rate is written as,
\begin{equation}
	\dot{m}_{pg,s} = \sum_{i \in [1,N_p]} \sum_{j \in [1,P_i]} \theta_{i,j,k} \epsilon_{i,v}\rho_{i,v}F_{i,j}\frac{\partial \chi_{i,j}}{\partial t},
\end{equation}
where $\epsilon_{i,v}$ and $\rho_{i,v}$ are the volume fraction and the density of the virgin phase $i$. $F_{i,j}$ and $\chi_{i,j}$ are the mass fraction and the advancement(0 - 1) of the sub-phase $j$, respectively. The $\chi_{i,j}$ is modelled using the Arrhenius law~\cite{TORRESHERRADOR2020105784},
\begin{equation}
	\frac{\partial \chi_{i,j}}{\partial t} = (1-\chi_{i,j})^{m_{i,j}} A_{i,j} \exp{(-\frac{E_{i,j}}{RT})}.
\end{equation}
where the rate parameters are given in Table~\ref{Tab.PICAparameters}. The overall production rate is obtained by summing over $ns$ reactions, 
\begin{equation}
	\dot{m}_{pg} = \sum_{s \in[1,ns]}\dot{m}_{pg,s} = -\frac{\partial (\epsilon \rho)_{pm}}{\partial t}.
\end{equation}
% \begin{table}[htb!]
% 	\centering
% 	\caption{Pyrolysis reactions for PICA.~\cite{MEURISSE2018497}}
% 	\label{Tab.sublimation}%添加标题 设置标签
% 	\begin{tabular}{p{4.5cm}cccccc}
% 		\toprule
% 		Pyrolysis reaction	&$F_{i,j}$	&$A_{i,j}\mathrm{(cm^3/mol-s)}$	&$E_{i,j}\mathrm{(J/mol)}$&	$m_{i,j}$	&$n_{i,j}$\\
% 		\midrule
% 		$pm_{2,1} \rightarrow \mathrm{0.69H_2O} + 0.02 \mathrm{C_6H_6} + 0.29 \mathrm{C_6H_5OH}$	&0.25	&1.2E4	&7.11E4	&3	&0 \\
% 		$pm_{2,2} \rightarrow \mathrm{0.09CO_2} + 0.33 \mathrm{CO} + 0.58 \mathrm{CH_4}$	&0.19	&4.98E8	&1.7E5	&3	&0 \\
% 		$pm_{2,3} \rightarrow \mathrm{H_2}$	&0.06	&4.98E8	&1.7E5	&3	&0 \\
% 		%\hline
% 		\bottomrule
% 	\end{tabular}
% \end{table}

\begin{table}[htb!]
	\centering
	\caption{Pyrolysis reaction parameters of the PICA.~\cite{MEURISSE2018497}}
	\label{Tab.PICAparameters}%添加标题 设置标签
	\begin{tabular}{p{7cm}ccccc}
		\toprule
		Formula	&$F_{i,j}$	&$A_{i,j}\mathrm{(\frac{cm^3}{mol-s})}$	&$E_{i,j}\mathrm{(\frac{J}{mol})}$&	$m_{i,j}$\\
		\midrule
		$pm_{2,1} \rightarrow \mathrm{0.69H_2O} + 0.02 \mathrm{A1} + 0.29 \mathrm{A1OH}$	&0.25	&1.2E4	&7.11E4	&3\\
		$pm_{2,2} \rightarrow \mathrm{0.09CO_2} + 0.33 \mathrm{CO} + 0.58 \mathrm{CH_4}$	&0.19	&4.98E8	&1.7E5	&3\\
		$pm_{2,3} \rightarrow \mathrm{H_2}$	&0.06	&4.98E8	&1.7E5	&3\\
		%\hline
		\bottomrule
	\end{tabular}
\end{table}

\subsubsection{Momentum conservation equation}
The flow of pyrolysis gases within them is described by Darcy's Law~\cite{ENE1982623},
\begin{equation}
	\label{Eq.darcy law}
	\bm{u} = -\frac{K}{\epsilon \mu}\nabla p,
\end{equation}
where $K$ is the permeability and is assumed to be uniform. The material properties are interpolated by the volume fraction and temperature, i.e., $K = cK_{v} + (1-c)K_c$, $\epsilon = c\epsilon_{v} + (1-c)\epsilon_c$, $\lambda = c\lambda_{v} + (1-c)\lambda_c$. The material properties of the virgin and char layer are taken from Theoretical Ablative Composite for Open Testing (TACOT)~\cite{TACOTv3.0} and are summarized in Table~\ref{Tab.porous}.
\begin{table}[htb!]
	\centering
	\caption{Material properties of charring ablative materials from TACOT~\cite{TACOTv3.0}.}
	\label{Tab.porous}%添加标题 设置标签
	\begin{tabular}{cccc}
		\toprule
		%\hline
		$K_v[\rm{m^2}]$&$K_c[\rm{m^2}]$&$\epsilon_v$&$\epsilon_c$\\
		\midrule
		$1.6e-11$&$2e-11$&0.8&0.85\\
		%\hline
		\bottomrule
	\end{tabular}
\end{table}

Introducing the equation of state $p_s = \rho_{s} RT/M_s$ and the Darcy's law (Eq.\ref{Eq.darcy law}) to the mass conservation equation (Eq.\ref{Eq.mass}) yields the momentum equation in terms of pressure,
\begin{equation}
	\label{Eq.pressure}
	\frac{\partial p_s}{\partial t} = \frac{p_s K}{\epsilon_{pg} \mu} \nabla^2 p + \frac{K}{\epsilon_{pg} \mu} (\nabla p_s \cdot \nabla p) + \nabla \cdot (p D_s \nabla Y_s) - \frac{RT}{\epsilon_{pg} M_s} \epsilon_{pm} \dot{m}_{s}, 
\end{equation}
where $p_s$ is the partial pressure, $\nabla^2 = \nabla \cdot \nabla$ is the Laplace operator, $\dot{m}_s=\dot{m}_{pg,s}+\dot{m}_{c,s}$ is the mass production rate of the $s$-th species, $\dot{m}_{c,s}$ is the chemical reaction rate of $s$-th species where the chemical reaction rate parameters are taken from Ref.~\cite{doi:10.2514/3.6330}. Solving pressure in the momentum equation offers advantages in handling the boundary condition. The multicomponent gas model solves the conservation equation of the species mass to consider the complexity of the component of the pyrolysis gases. The species diffusion and finite rate chemistry of the pyrolysis gases are taken into account. If the pyrolysis gas is assumed to be mixture gases, the momentum equation is simplified as,
\begin{equation}
	\label{Eq.pressure2}
	\frac{\partial p}{\partial t} = \frac{p K}{\epsilon_{pg} \mu} \nabla^2 p + \frac{K}{\epsilon_{pg} \mu} (\nabla p \cdot \nabla p) - \frac{RT}{\epsilon_{pg} M} \epsilon_{pm} \dot{m}_{pg}
\end{equation}

The mix gas model calculates thermodynamic and transport parameters of the pyrolysis gases by looking up a table of experimentally calibrated parameters. Compared with the mixed gas model, the multicomponent gas model calculates the thermophysical and transport parameters by the Wilke's mixing law~\cite{gupta1990review} since the species concentration is solved. Thus, the pre-experimentally calibrated parameter tables of the thermophysical and transport parameters are no longer required for the multicomponent gas model.

\subsubsection{Energy conservation equation}
The conservation equation of energy is written as,
\begin{equation}
		\label{Eq.energy}
	\frac{\partial (\epsilon \rho e)_{pm}}{\partial t} + \frac{\partial(\epsilon \rho e)_{pg}}{\partial t} + \nabla \cdot (\epsilon \rho h \bm{u})_{pg} = \nabla \cdot (\lambda \nabla T).
\end{equation}
Assuming local thermal equilibrium, the pyrolysis gas and polymer matrix use a unified temperature $T$ to represent the internal energy, i.e., $h = c_{p} T,  e = c_{v} T$. Substituting $\kappa = \rho_{pm}\epsilon_{pm}c_{v,pm}+\rho_{pg}\epsilon_{pg}c_{v,pg}$ into Eq.~\ref{Eq.energy} yields,
\begin{equation}
	\begin{split}
	\kappa \frac{\partial T}{\partial t} + \nabla \cdot (\epsilon \rho h \bm{u})_{pg} = -\frac{\partial (\epsilon \rho)_{pm}}{\partial t} h_{pm} - \frac{\partial (\epsilon \rho)_{pg}}{\partial t} h_{pg}
 + \frac{\epsilon_{pg} \partial p}{\partial t} + \nabla \cdot (\lambda \nabla T).
	\end{split}
\end{equation}

\subsubsection{Implicit time integration in three dimensional curvilinear system}

The governing equations of the MTR are transformed into the generalized space $(\xi,\eta,\zeta)$ to accommodate arbitrary curvilinear coordinates of body-fitted grids. The governing equations are written in dual time-step form,
\begin{equation}
	\label{Eq.iterate}
	\frac{1}{J}\frac{\partial \bm{W}}{\partial \tau} + \frac{1}{J}\frac{\partial \bm{W}}{\partial t} +\bm{R}(\bm{W})= 0,
\end{equation}
where $\bm{W} = [p_1, \cdots, p_{ns},\kappa T]$ is the vector of the solution variables, $J$ is the determinant of the curvilinear coordinate transformation Jacobian, $\bm{R}$ is the spatial residual and is given in detail in the Appendix A. $\tau$ is the pseudo iteration. When the pseudo iteration convergences $\partial \bm{R}/\partial \tau = 0$, the governing equations degenerate to the original form, $\partial \bm{W}/\partial t+J\bm{R}=0$.

The differential form of the above equation in the implicit time level is written as, 
\begin{equation}
	\label{Eq.iterate}
	\begin{array}{l}
	(\frac{1}{J \Delta \tau}) \Delta \bm{W}^{m+1} = 
	-\bm{R}(\bm{W}^{m+1}) - \frac{(1+\theta)(\bm{W}^{m+1}-\bm{W}^n) - \theta(\bm{W}^n - \bm{W}^{n-1})}{J \Delta t},
	\end{array}
\end{equation}
where $\Delta \bm{W}^{m+1} = \bm{W}^{m+1} - \bm{W}^m$ is the increment of the conservative variables, superscript $m, n$ denote the pseudo time level and the physical time level, respectively. The parameter $\theta$ is the time differential coefficient with $\theta = 0.5$ for second-order time discretization and $\theta = 0$ for first-order time discretization. $\Delta \tau$ is computed by Courant-Friedrichs-Lewy (CFL) condition, 

\begin{equation}
	\Delta \tau = \frac{CFL/J}{\Lambda^{\xi}+\Lambda^{\eta}+\Lambda^{\zeta}},
\end{equation}
where $CFL$ and $\Lambda$ are the CFL number and the spectral radii of flux jacobian, respectively. The spectral radius of the MTR solver is calculated by,
% \begin{equation}
% 	\Lambda_c^{\xi} = \left| U \right| + c, \Lambda_v^{\xi} = \max{(\frac{4}{3},\frac{\gamma}{\rho})(\frac{\mu_L}{Pr_L}+\frac{\mu_T}{Pr_T})\frac{(\Delta S^{\xi})^2}{\Omega}},
% \end{equation}
% where $U=u\xi_x+v\xi_y+w\xi_z$ is the contravariant velocity, $Pr$ is the Prandtl number, subscripts $L, T$ denote the laminar and the turbulent, $\Delta S$ is the face area of the control volume. For Eq.\ref{Eq.pressure},
\begin{equation}
	\begin{array}{c}
	\Lambda^{\xi}_p = \frac{p_s}{\epsilon_{pg} \mu}(\xi_x^2+\xi_y^2+\xi_z^2),\\
	\Lambda^{\xi}_T = (\epsilon \rho c_{v} U)_{pg} + \frac{\xi_s \rho_s c_{v,s}}{\lambda}(\xi_x^2+\xi_y^2+\xi_z^2).
	\end{array}
\end{equation}
Similar expressions hold for the $\eta$ and $\zeta$ direction. The residual $\bm{R}$ is linearized by the Taylor expansion, 

\begin{equation}
  \label{Eq.linear}
  \bm{R(W)}^{m+1} = \bm{R(W)}^m + \frac{\partial \bm{R}}{\partial \bm{W}}|^m \Delta \bm{W}^{m+1} + O(\left\| \Delta \bm{W}^{m+1}\right\|)^2,
\end{equation}
where ${\partial \bm{R}}/{\partial \bm{W}}$ is the flux jacobian. The flux Jacobian of the MTR solver is provided upon Appendix A. The high-order truncation term $O(\Delta \bm{W}^{m+1})$ in Eq.(\ref{Eq.linear}) is discarded. Hence, the linearized implicit iteration scheme of Eq.(\ref{Eq.iterate}) is reformulated as 
\begin{equation}
	\begin{array}{l}
  \label{Eq.linearizedimplicit}
  (\frac{1}{J \Delta \tau}\bm{I} + \frac{1+ \theta}{J \Delta t} \bm{I}+ \frac{\partial \bm{R}}{\partial \bm{W}}) \Delta \bm{W}^{m+1} =\\ -\bm{R}(\bm{W}^{m}) - \frac{(1+\theta)(\bm{W}^m-\bm{W}^n) - \theta(\bm{W}^n - \bm{W}^{n-1})}{J \Delta t},
	\end{array}
\end{equation}
where $\bm{I}$ is the identity matrix. The above equation can be solved by the LUSGS~\cite{doi:10.2514/3.10007} method. The detailed derivation of the implicit time integration for the governing equations of hypersonic flows can refer to Ref.~\cite{10.1063/5.0218188}.

\begin{figure}[htb!]
	\centering
	\subfigure
	{
		\includegraphics[width=0.46\textwidth]{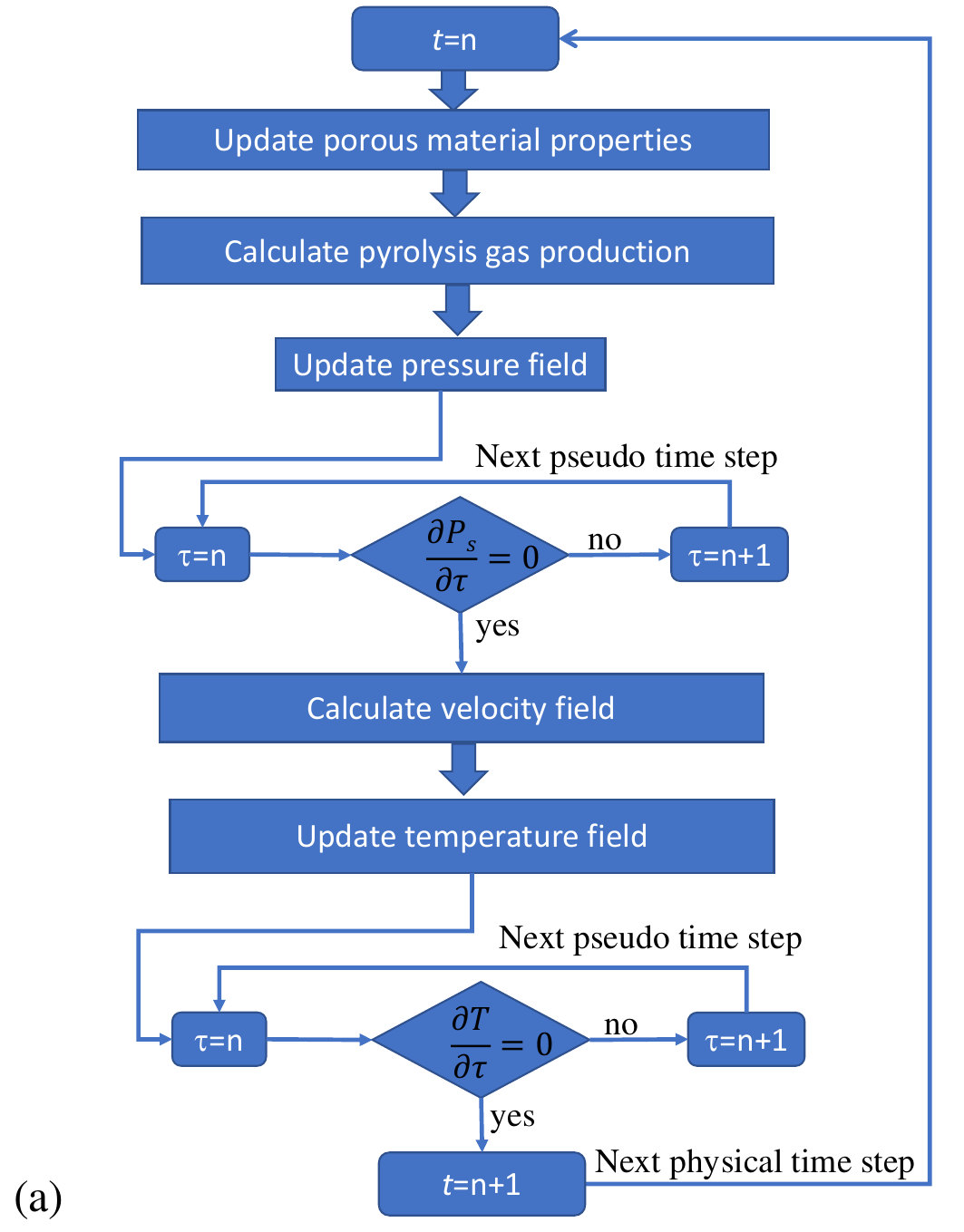}
	}	
	\subfigure
	{
		\includegraphics[width=0.46\textwidth]{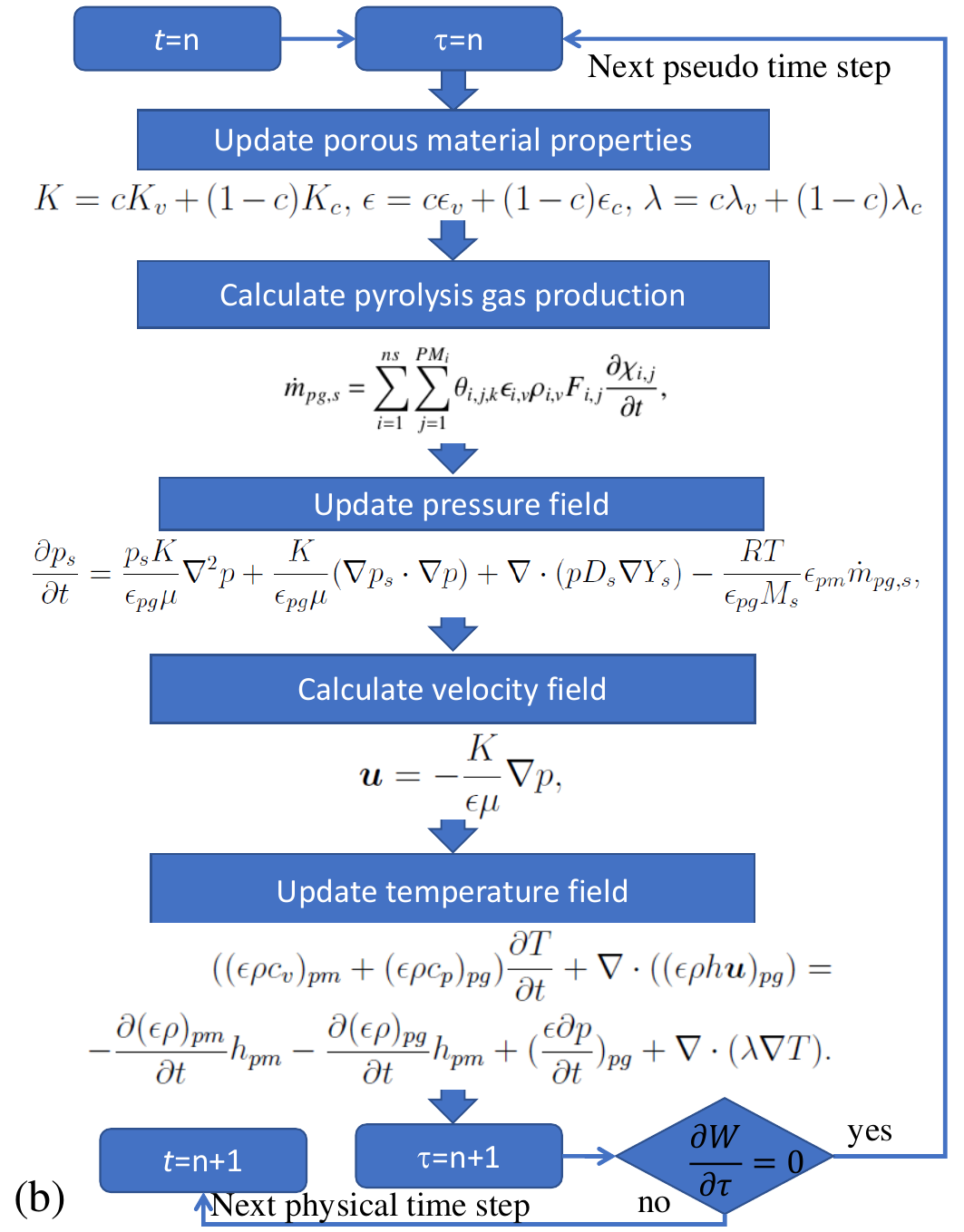}
	}
		% \centering
		% \includegraphics[width=0.6\textwidth]{figure/coupling.pdf}
	\caption{The schematic of the coupling solution of the MTR:(a)explicit coupling solution;(b)implicit coupling solution.}
	\label{Fig.MTRsolve}
\end{figure}

The conservation equations of the pressure and temperature can be solved by an explicit coupling solution or an implicit coupling solution. The schematic of these coupling solutions is illustrated in Fig.~\ref{Fig.MTRsolve}. The explicit mechanism iteratively and independently solves the pressure and temperature, freezing the unsolved variables during the pseudo iteration. The conservation equations do not need to reach convergence at the same pseudo time step. Thus, the computational efficiency is more efficient. In contrast, the implicit coupling mechanism updates the pressure and temperature in every pseudo iterations. The conservation equations need to reach convergence at the same pseudo time step. Thus, the time discretization error is reduced.

\subsection{Gas-surface interaction}%The governing equations of the porous material response are solved by our in-house code with serval validation cases, which is embedded with a customizable gas-surface interaction (GSI) interface to couple with our chemically reacting flow solver to simulate the unsteady multi-physics of TPS materials. 
\begin{figure}[htb!]
	\centering
		\includegraphics[width=0.8\textwidth]{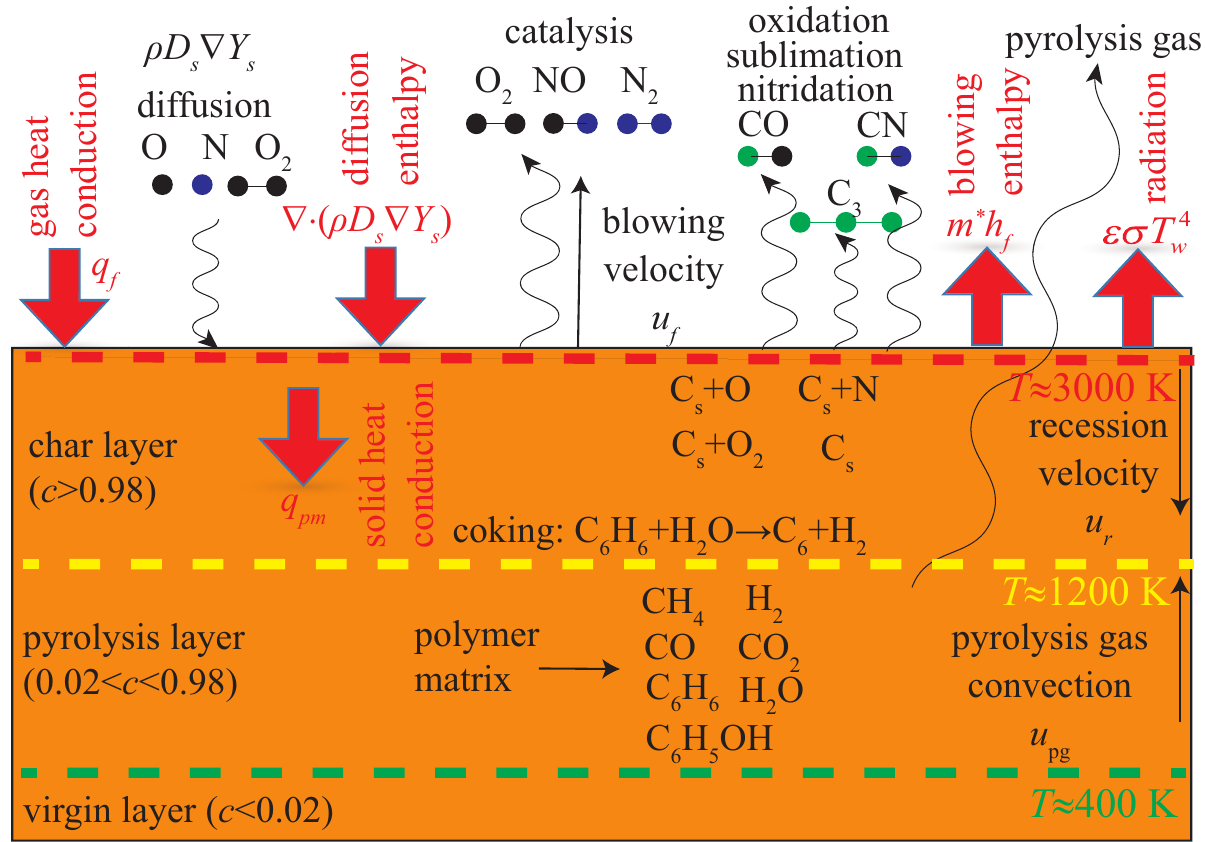}
	\caption{Schematic of the gas-surface interaction with the SMB and SEB.}
	\label{Fig.GSI}
\end{figure}

The solutions of CRF and MTR are coupled by a customizable gas-surface interaction interface. Assuming a thin-film on the surface, the mass and energy transfer within this thin-film are explained through the surface mass balance (SMB) and surface energy balance (SEB), as illustrated in Fig.~\ref{Fig.GSI}. The SMB accounts for the conservation equation of mass that the mass flux entering the thin-film due to diffusion and surface reactions must be equal to the mass flux exiting the thin-film due to convection (blowing),
\begin{equation}
	\label{Eq.SMB}
	\nabla \cdot (\rho D_s \nabla Y_s)_f + \dot{m}_{a,s} + (\rho_{s}u)_{pg} = (\rho u Y_s)_w,
\end{equation}
where $\dot{m}_{a,k}$ is the gas production rate from ablation. Summing Eq.\ref{Eq.SMB} over all species generates the blowing velocity $u_{w}$ and the recession velocity $r$,
\begin{equation}
	(\rho u)_{w} = \sum_{s=1}^{ns}(\dot{m}_{a,k} + (\rho_{s}u)_{pg} ) = \rho_{pm} \cdot r.
\end{equation}

The SEB describes the energy flux entering the thin-film must be equal to the energy flux leaving the thin-film, 
\begin{equation}
	\begin{split}
	q_f + \sum_{s=1}^{ns} (\rho h \nabla \cdot (D_s \nabla Y_s))_{f} + \epsilon \sigma(T_{sur}^4-T_w^4) + \dot{m}_a h_{w} \\ + \sum_{s=1}^{ns}(\rho_s u h_s)_{pg} = (\rho u h)_w +q_{pg} + q_{pm},
	\end{split}
\end{equation}
where $q_f, q_{pg}, q_{pm}$ are the heat conduction of external flows, pyrolysis gases, and polymer matrix, respectively. The second term is the gaseous enthalpy carried by diffusion. The third term is radiation heat. The 4-th term is the enthalpy change of surface ablation. The 5-th term is the convection enthalpy of pyrolysis gases. The 6-th term is the convection enthalpy of surface blowing. The surface reaction rate is derived based on the kinetic theory,

\begin{equation}
	\label{Eq.rate}
	k_{m} = \alpha_m\sqrt{\frac{RT_w}{2\pi M_s}},
\end{equation}
where $\alpha_{m}$ is the reaction probability for reaction $m$, $R$ is the universal gas constant, $M_s$ is the molecular mass of the $s$-th species. The surface reaction chemistry considers oxidation, nitridation, sublimation, and catalyst with the reaction parameters are summarized in Table~\ref{Tab.oxidation}. 

\begin{table}[htb!]
	\centering
	\caption{Reaction rate parameters of surface chemistry.}
	\label{Tab.oxidation}%添加标题 设置标签
	\begin{tabular}{ccc}
		\toprule
		Ref.&formulation	&reaction probability\\
		\midrule
		\multirow{4}*{Park.~\cite{doi:10.2514/2.6426}}&
		$\rm{C_s + O \rightarrow CO}$&$0.63\exp(-1160/T_w)$\\
		% $\rm{3C_s \Leftrightarrow C_3}$&$0.1\exp(-5.19 \times 10^{15}/R T_w)$\\
		~&$\rm{2C_s + O_2 \rightarrow 2CO}$&$0.5$\\
		~&$\rm{C_s + N \rightarrow CN}$&$0.3$\\
		~&$\mathrm{C_s + O_2 \rightarrow CO + O}$	&$\frac{1.43 \times 10^{-3} + 0.01\mathrm{exp}(-1450/T_w)}{1 + 2 \times 10^{-3}\mathrm{exp}(13000/T_w)}$\\
		~&$\rm{3C_s \rightarrow C_3}$&$5.19 \times 10^{13} \exp(-93312/T_w)$\\
		\midrule
		\multirow{3}*{Driver.~\cite{doi:10.2514/6.2011-3784}}&
		$\rm{C_s + O \rightarrow CO}$&$0.9$\\
		~&$\rm{2C_s + O_2 \rightarrow 2CO}$&$0.01$\\
		~&$\rm{C_s + N \rightarrow CN}$&$0.005$\\
		%\hline
		\midrule
		\multirow{3}*{Catalyse.~\cite{doi:10.2514/6.2011-3784}}&
		$\rm{N + N\Leftrightarrow N_2}$&$0.001$\\
		~&$\rm{O + O\Leftrightarrow O_2}$&$0.001$\\
		~&$\rm{N + O\Leftrightarrow NO}$&$0.001$\\
		\bottomrule
	\end{tabular}
\end{table}

Sublimation is calculated by the Knudsen-Langmuir equation~\cite{doi:10.2514/3.60806},

\begin{equation}
	\label{Eq.sublimation}
	\dot{m_s} = \alpha_s(p_{\rm{eq}}-p_s)\sqrt{\frac{M_s}{2\pi RT_w}},
\end{equation}
where $\alpha_s$ is a experimental calibration parameter, $p_{\rm{eq}}$ is the equilibrium vapor pressure, and $p_s$ is the actual vapor pressure. These parameters are summarized in Table~\ref{Tab.sublimation}.%which obtained from Ref.\cite{doi:10.2514/6.1968-754}
\begin{table}[htb!]
	\centering
	\caption{Sublimation reaction parameters.}
	\label{Tab.sublimation}%添加标题 设置标签
	\begin{tabular}{cccc}
		\toprule
		Ref.&formulation	&$\alpha_s$&$p_{\rm{eq}}$\\
		\midrule
		\multirow{3}*{Mortensen.~\cite{doi:10.2514/1.J052659}}&
		$\rm{C_s \Leftrightarrow C}$&0.14&$\exp(-85715/T_w+18.69)$\\
		~&$\rm{2C_s \Leftrightarrow C_2}$&0.26&$\exp(-98363/T_w+22.20)$\\
		~&$\rm{3C_s \Leftrightarrow C_3}$&0.03&$\exp(-93227/T_w+23.93)$\\
		%\hline
		\bottomrule
	\end{tabular}
\end{table}

\subsection{Simplified ablation boundary condition}
In the coupled simulation between CRF and MTR, the time-consuming MTR solutions can be approximated by a quasi-steady ablation boundary to achieve higher efficiency. When the hypersonic flows have been solved in the CRF solver, the unclosed terms in the SEB is the solid heat conduction, $q_{cond} = q_{pm} + q_{pg}$. The quasi-steady ablation boundary introduce serval simplification assumptions to approximate the $q_{cond}$ and avoid time-consuming MTR solutions. Thus, the coupled simulation only requires CRF solver with an quasi-steady ablation boundary. One such simplification is the radiation equilibrium (RE) assumption which assumes that the surface reaches thermal equilibrium without solid heat conduction ($q_{cond} =0$). Another simplification is the steady-state ablation (SSA) assumption, which considers the solid phase as a semi-infinite material and approximates the solid heat conduction with a function related to the surface enthalpy ($q_{cond} = \dot{m}_{a}(h_w - h_v)$). While, $h_w$ is calculated according to the temperature, and $\dot{m}_a$ is calculated by the surface reaction rate. These assumptions build an simplified ablation boundary condition to calculate the solid heat conduction and solve the ablative flow fields in a steady or quasi-steady state without coupling the MTR solutions. This ablation boundary is described as follows:
\begin{enumerate}
	\item The partial density $\rho_{f,s}$, and the temperature $T_w$ on the ablating surface are used to compute the diffusion velocity gradient $\nabla{Y_s}$ and the temperature gradient $\nabla{T}$ through the SMB and the SEB, respectively.
	\item The $\nabla{Y_s}$ and $\nabla{T}$ are enforced in the boundary by extrapolating the partial density and temperature of the ghost cells. 
	\item The pressure gradient and the velocity on the ablation boundary are set to be 0 and the blowing velocity $u_f$, respectively.
\end{enumerate}
 
% , with less compromising of accuracy
\subsection{Fully implicit coupling method}
The fully implicit coupling mechanism solves the governing equations of the CRF and MTR in three-dimensional curvilinear coordinate $(\xi,\eta,\zeta)$ with a dual time-step~\cite{jameson1991time} form, 
\begin{equation}
	\label{Eq.implicit}
	\frac{1}{J}\frac{\partial \bm{Q}}{\partial\tau} + \frac{1}{J}\frac{\partial \bm{Q}}{\partial t} + \bm{R}(\bm{Q}) = 0
\end{equation}%where $\bm{Q}$ is the vector of $\bf{U}$ in Eq.~\ref{Eq:govern},  in Eq.\ref{Eq.pressure}, and temperature in Eq.~\ref{Eq.energy},
where $\bm{Q} = [\bm{U}, \bm{W}]^{\mathrm{T}}$ is the the vector of solution variables of the CRF and MTR, and $\bm{R}$ is the residual. The differential form of Eq.\ref{Eq.implicit} is similar to Eq.\ref{Eq.linearizedimplicit}, while the cross term of the flux jacobian between CRF and MTR is $\bm{0}$. 
% \begin{equation}
% 	\label{Eq.timediscretisation}
% 	(\frac{\bm{Q}^{m+1}-\bm{Q}^{m}} {J \tau} + \frac{(1+\Phi)(\bm{Q}^{m+1}-\bm{Q}^n) - \Phi(\bm{Q}^n - \bm{Q}^{n-1})}{J t} ) = \bm{R}(\bm{Q}^{m+1}),
% \end{equation}
% where $\Phi = 0.5$. $m, n$ is the pseudo time level and physical time level, respectively. The residual is linearised in pseudo time as follows, 
% \begin{equation}
% \bm{R}(\bm{Q}^{m+1}) = \bm{R}(\bm{Q}^{m}) + \frac{\partial \bm{R}}{\partial \bm{Q}} \Delta \bm{Q}^m.
% \end{equation}
% \begin{equation}
% 	\bm{RHS} = \bm{R(\bm{Q}^m)} - \frac{(1+\Phi)(\bm{Q}^{m+1}-\bm{Q}^n) - \Phi(\bm{Q}^n - \bm{Q}^{n-1})}{J t}.
% \end{equation}

% The differential form of Eq.(\ref{Eq.linearizedimplicit}) is written as, 
% \begin{equation}
% 	\label{Eq.implicitscheme}
%   \begin{array}{l}
%     \left\{   \bm{I}_m + J^{-1}\Delta \tau  [\partial_{\xi} (\bm{A}_{\xi,inv} + \bm{A}_{\xi,vis}) + \partial_{\eta} (\bm{A}_{\eta,inv} + \bm{A}_{\eta,vis}) + \partial_{\zeta} (\bm{A}_{\zeta,inv} + \bm{A}_{\zeta,vis}) ] \right\} \\ \Delta \bm{Q}^{m+1} = \bm{LHS} \Delta \bm{Q}^{m+1}  = \bm{RHS}^m,
%   \end{array}
% \end{equation}
% where $\partial_\xi, \partial_\eta, \partial_\zeta$ is the differential operator in the $\xi,\eta,\zeta$ direction. $\bm{LHS}$ and $\bm{RHS}$ refer to the left-hand side (implicit operator) and the right-hand side that determine the convergence characteristics and the spatial residuals, respectively. 

\begin{figure}[htb!]
	\centering
	\subfigure
	{
		\includegraphics[height=0.38\textwidth]{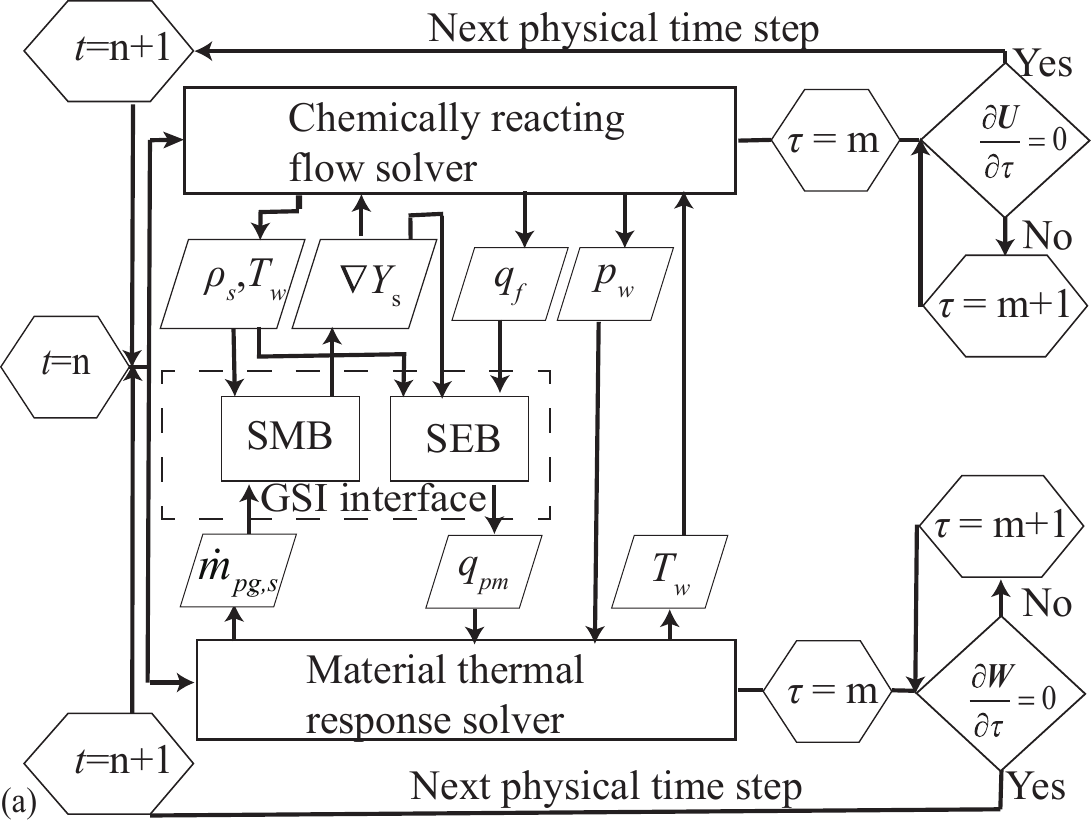}
	}	
	\subfigure
	{
		\includegraphics[height=0.38\textwidth]{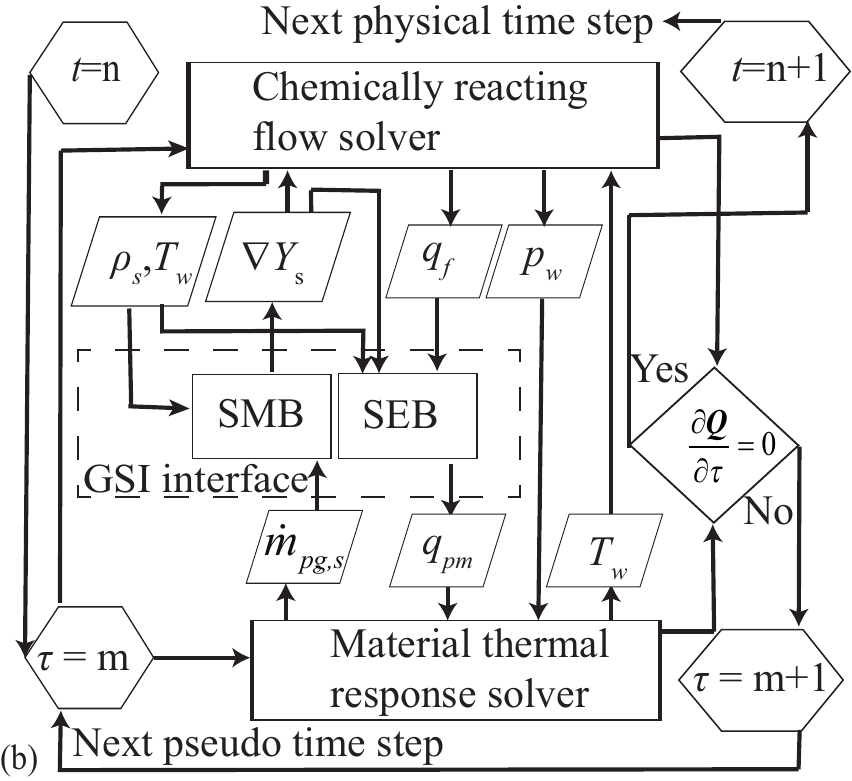}
	}
	\caption{Illustration of the gas-surface interaction mechanism between chemical reacting flows and material thermal response:(a)explicit coupling;(b)implicit coupling.}
	\label{Fig.coupling}
\end{figure}

The interfacial quantities on the ablating wall are modelled by the proposed GSI interface. Depending on when the GSI interface is called to update the interfacial quantities, the solutions are classified as explicit coupling mechanism and implicit coupling mechanism, as shown in Fig.~\ref{Fig.coupling}. Traditional method usually employs explicit coupling method that the physical quantities on the surface between CRF and MTR are exchanged at physical time level. Then, the ablation boundary conditions of the CRF and MTR solvers are freezed during the pseudo iteration after the GSI interface is called. However, this explicit coupling approach introduces temporal discretization errors and numerical oscillations due to the abrupt change of the interfacial quantities. In contrast, the fully implicit coupling method update interfacial quantities at the begin of the pseudo iteration, ensuring temporal discretization accuracy. The pseudo iteration is convergent when the unsteady residual $\partial \bm{Q}/\partial \tau$ achieves a predefined tolerance for both the CRF and MTR solvers. Thus, the time discretization error is eliminated since the interfacial quantities are updated in the pseudo iteration. Meanwhile, the interfacial quantities are continuously changed without numerical oscillations.

The GSI interface is the footstone of the coupled simulation between the CRF solver and the MTR solver. First, the partial density $\rho_{f,s}$, the pyrolysis production rate $\dot{m}_{pg,s}$, and the temperature $T_w$ on the surface are used to compute the diffusion velocity gradient $\nabla{Y_s}$ and the solid heat conduction $q_{cond}$ through the SMB and the SEB, respectively. Second, the concentration gradient $\nabla{Y_s}$ and blowing velocity $u$ are set as the boundary condition of the CRF solver on the interface. Finally, $q_{cond}$ provides a heat flux boundary for the MTR solver to calculate the variant of the surface temperature. The proposed GSI model employs the SMB and SEB equations to account for the mass and energy transfer on the ablating surface, which achieves strong coupling between the CRF and MTR solvers. It should be noted that the exchanging process of the interfacial quantities between the CRF and MTR solvers are implemented by the Message Passing Interface (MPI) technique.
 %comprised of physical variables and their gradients on the surface, 

\section{Numerical cases}
% 准备算例有1、IRV用于对比隐式耦合计算和传统耦合算法的对比，预期结果考虑壁面烧蚀效应后计算精度提高
% 2、通过辐射热平衡条件和定常烧蚀假设
% 2、UHTC
%飞行器烧蚀材料全弹道全耦合仿真
% Serval cases are conducted to show the

 \subsection{MTR test cases}

\begin{figure}[htb!]
	\centering
		\includegraphics[width=0.5\textwidth]{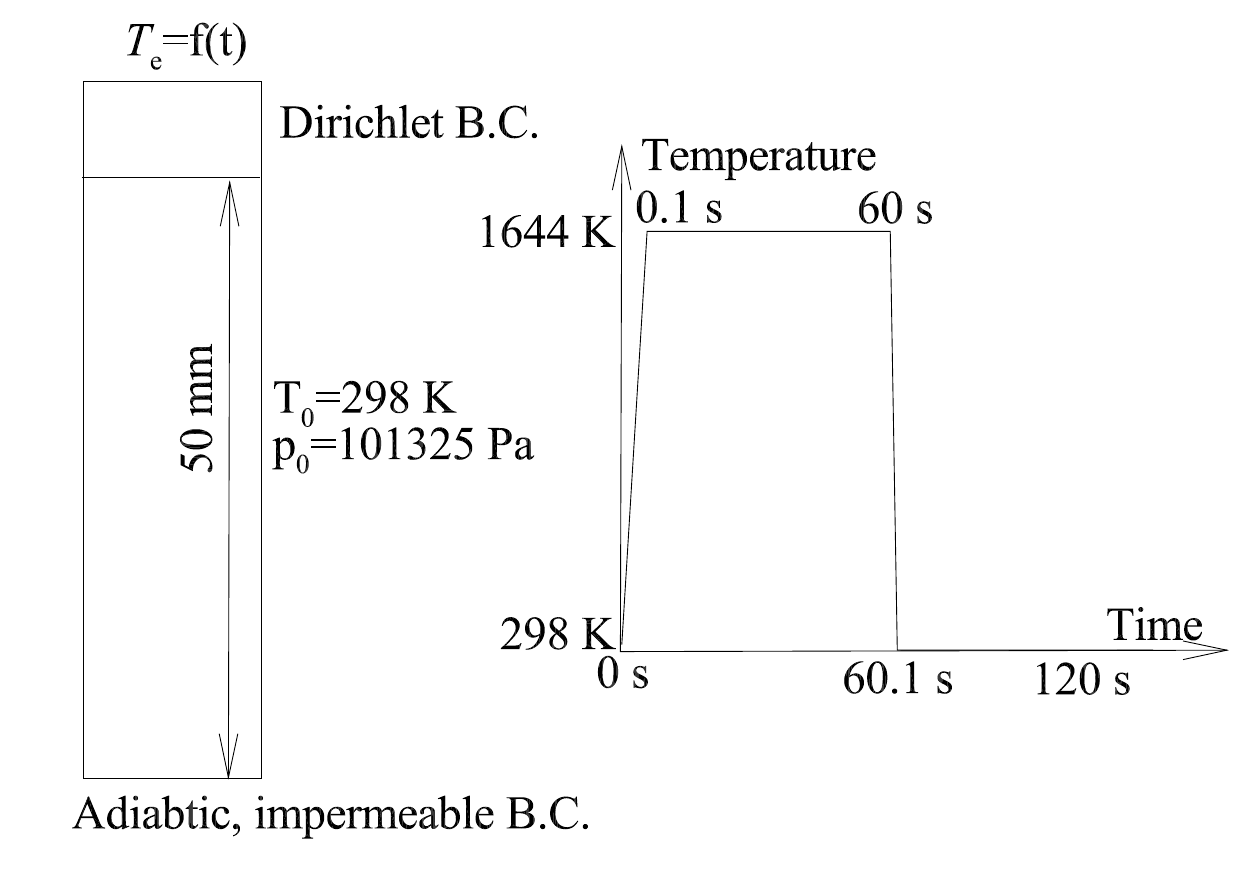}
		% \label{Fig.TACOTcase1}
	\caption{Schematic of the initial conditions, boundary conditions, and geometry of the MTR test case.}
	\label{Fig.TACOTcase}
\end{figure}

The code validation of the MTR uses standardized validation cases which are established by the Ablation Workshop in 2011~\cite{lachaud2011ablation}. The material properties and the numerical setup are detailed in Ref.~\cite{Lachaud2014Porous}. The validation case involves a one-dimensional convective heat transfer problem in a 50mm thick sample of charring ablative materials. One side of the sample is subjected to constant temperature boundary at 1644 K under a pressure of 1 atm for 1 minute. The opposing side of the sample employs impermeable and adiabatic boundary conditions. The geometry, initial conditions, and boundary conditions of the MTR test case are shown in Fig.\ref{Fig.TACOTcase}.

\subsubsection{Comparison of explicit and implicit time integration.}

\begin{figure}[htb!]
	\centering
	\subfigure
	{
		\includegraphics[width=0.45\textwidth]{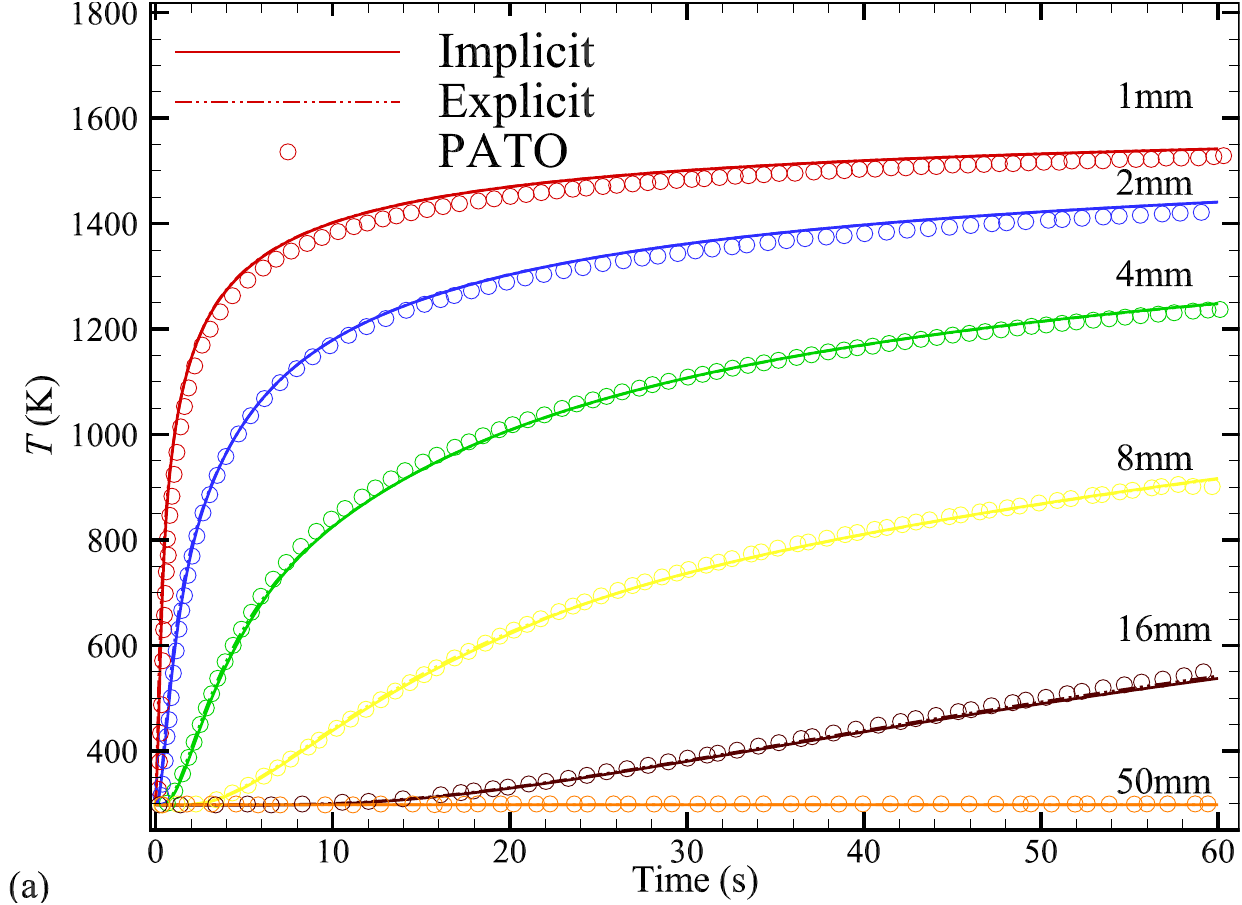}
	}	
	\subfigure
	{
		\includegraphics[width=0.45\textwidth]{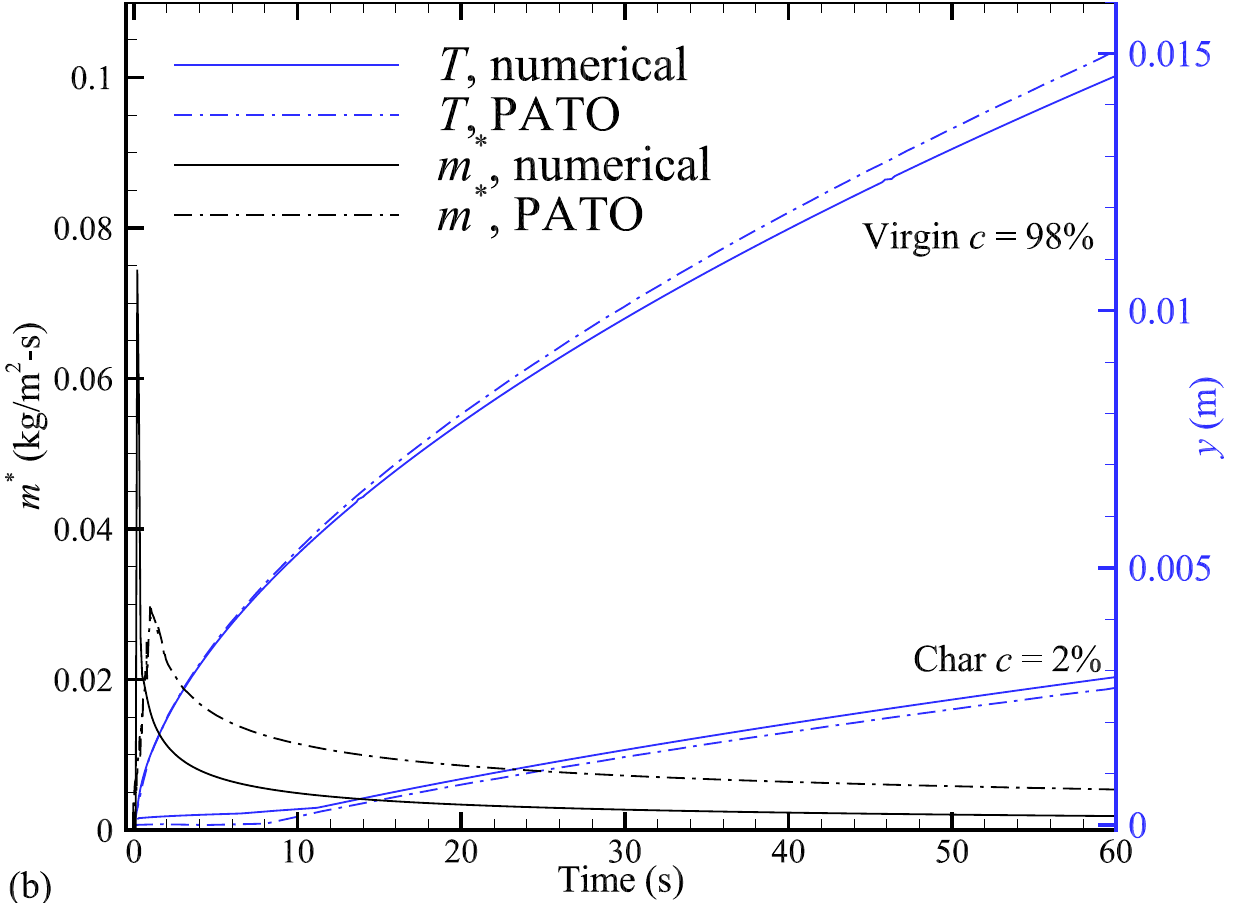}
	}	
	\caption{Comparison of the ablation quantities between explicit and implicit time integration in the TACOT case:(a)temperature at different profiles over time;(b)ablation rate, char layer position, and virgin layer position over time.}
	\label{Fig.TACOT}
\end{figure}

Traditional MTR method usually use explicit time integration, in which the maximum time step is limited by the CFL condition. As indicated in Table~\ref{Tab.dt}, the explicit time integration employs 3-step Runge-Kutta method~\cite{jameson1985numerical} with the maximum physical time step is limited to $10^{-8}$ s to maintain numerical stability. While, the implicit time integration allows for a larger physical time step of 5 s. Here, a time step of 0.1 s is used to reduce time discretization error and achieve higher accuracy. Compared to the explicit time integration, the computational time of the implicit time integration is reduced by 92\%. A comparison between the implicit and the explicit schemes is presented in Fig.~\ref{Fig.TACOT}. The results calculated by the explicit time integration and implicit time integration are in good agreement with the benchmarking calculation results from PATO~\cite{Lachaud2014Porous}, validating the accuracy of the current numerical methods in calculating the MTR of charring ablative materials. %The implicit method does not compromise accuracy because the implicit scheme agree well with the explicit method.

 \begin{table}[htb!]
	\centering
	\caption{Comparison of computing time between the explicit and implicit time integration of the MTR solutions.}
	\label{Tab.dt}%添加标题 设置标签
	\begin{tabular}{ccc}
		\toprule
		Method&Time step(s)&Total time(h)\\
		\midrule
		3-step Runge-Kutta (explicit)&1E-8&38.9\\
		LUSGS (implicit)&1E-1&2.98\\
		%\hline
		\bottomrule
	\end{tabular}
\end{table}

\subsubsection{Comparison of explicit and implicit coupling solutions of the MTR solver.}

\begin{figure}[htb!]
	\centering
	\subfigure
	{
		\includegraphics[width=0.45\textwidth]{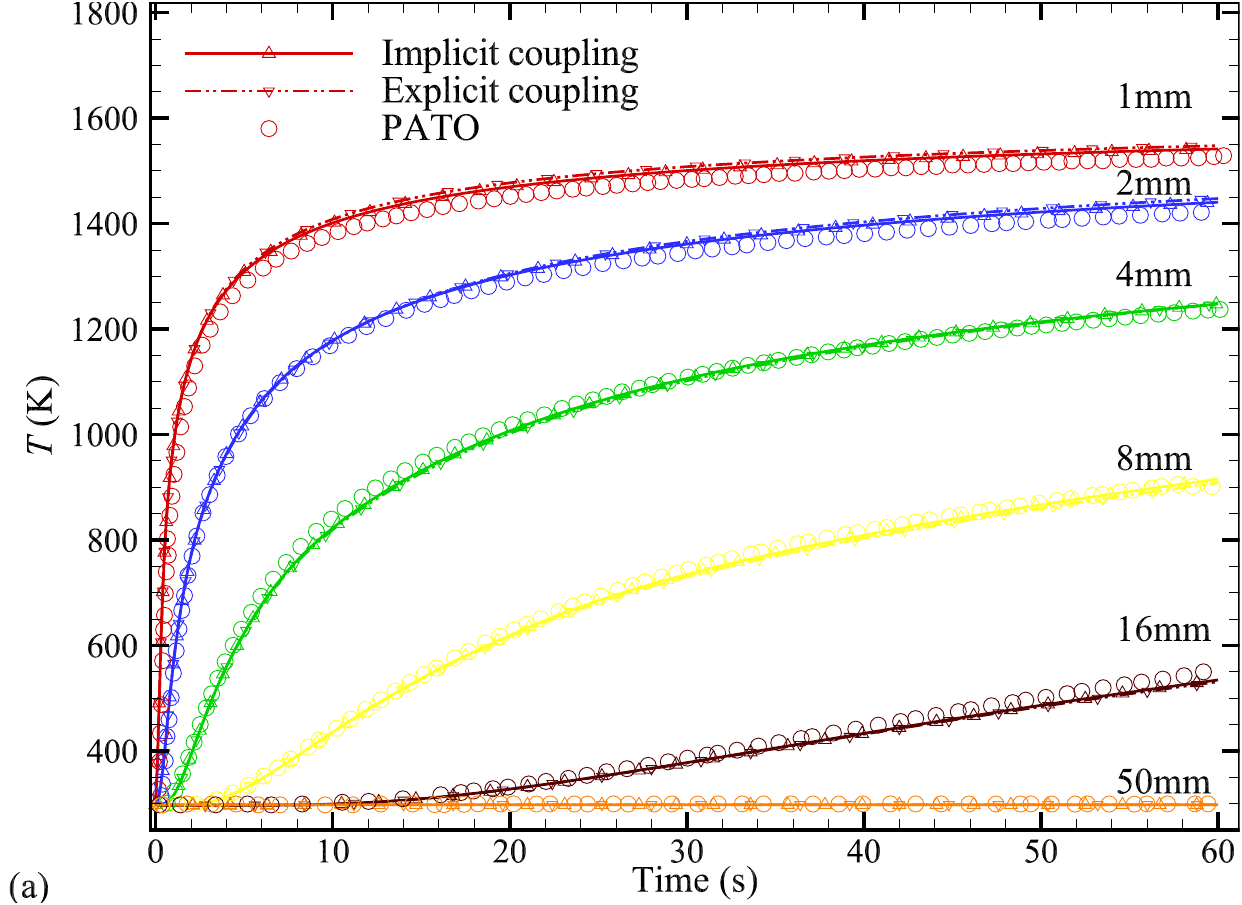}
	}	
	\subfigure
	{
		\includegraphics[width=0.45\textwidth]{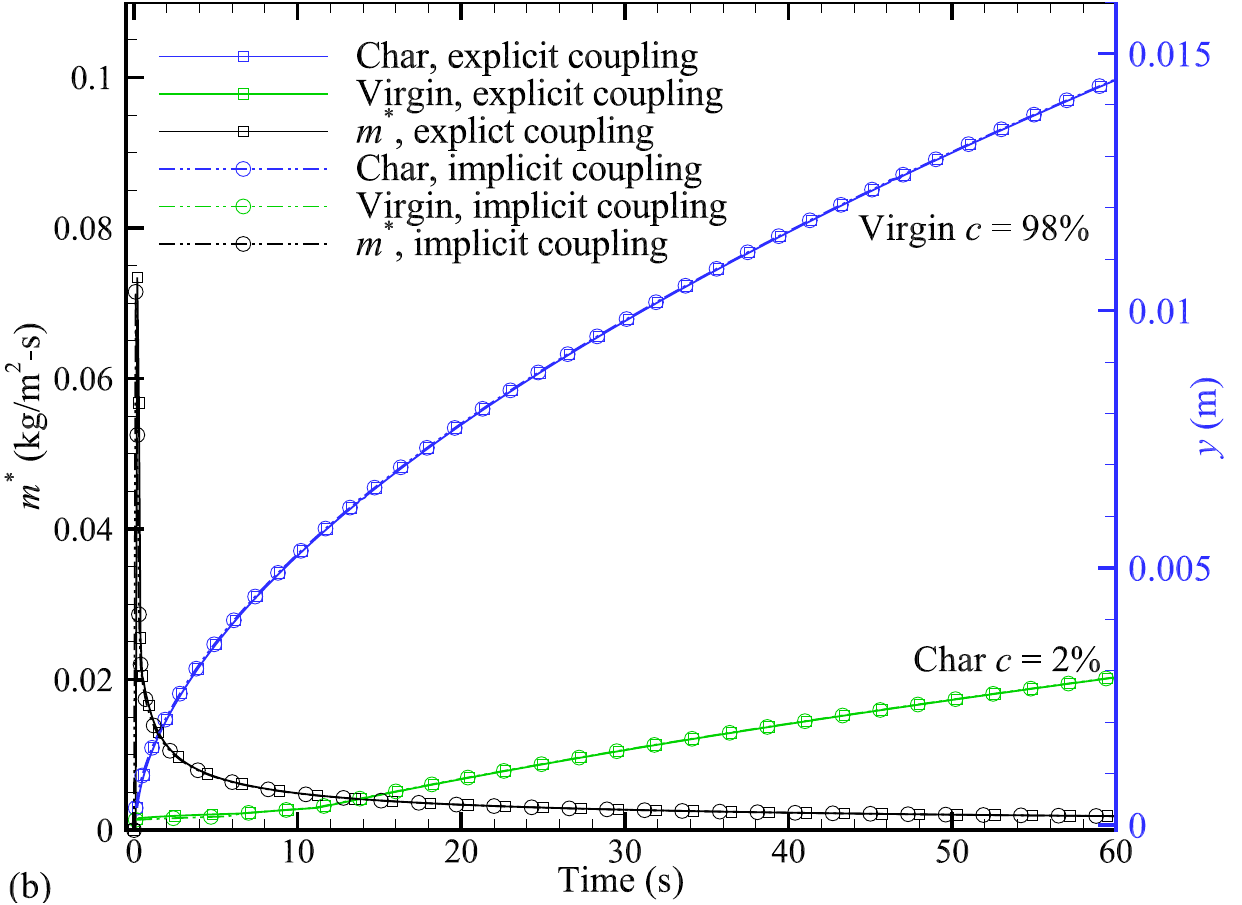}
	}	
	\caption{Comparison of the ablation quantities between explicit coupling and implicit coupling mechanisms in the TACOT case:(a)temperature at different profiles over time;(b)ablation rate, char layer position, and virgin layer position over time.}
	\label{Fig.TACOT-exp-imp}
\end{figure}

This case compares the accuracy and efficiency of the explicit and implicit coupling mechanism of the MTR solver, as illustrated in Fig.~\ref{Fig.MTRsolve}. The explicit coupling method separately solves the pressure and temperature at the physical time step. Thus, the variants of pressure and temperature are ignored during the pseudo time integration, resulting in time discretization errors. A comparison of the time consumption between the explicit coupling mechanism and the implicit coupling mechanism is given in Table~\ref{Tab.couplingdt}. The implicit coupling solution takes 4.76 times longer than the explicit coupling solution, indicting higher efficiency of the explicit coupling solution. Futhermore, the results calculated by these two coupling methods, as shown in Fig.~\ref{Fig.TACOT-exp-imp}, are essentially identical. In conclusion, the explicit coupling solution of the conservation equations of pressure and temperature does not compromise the accuracy but achieves higher efficiency, which is employed in the subsequent cases.
\begin{table}[htb!]
	\centering
	\caption{Comparison the time consumption between the explicit and implicit coupling solutions.}
	\label{Tab.couplingdt}%添加标题 设置标签
	\begin{tabular}{ccc}
		\toprule
		Method&Explicit coupling&Implicit coupling\\
		\midrule
		Time(h)&2.98&14.2\\
		%\hline
		\bottomrule
	\end{tabular}
\end{table}

\subsubsection{Comparison of multicomponent gas model and mix gas model}

\begin{figure}[htb!]
	\centering
	\subfigure
	{
		\includegraphics[width=0.45\textwidth]{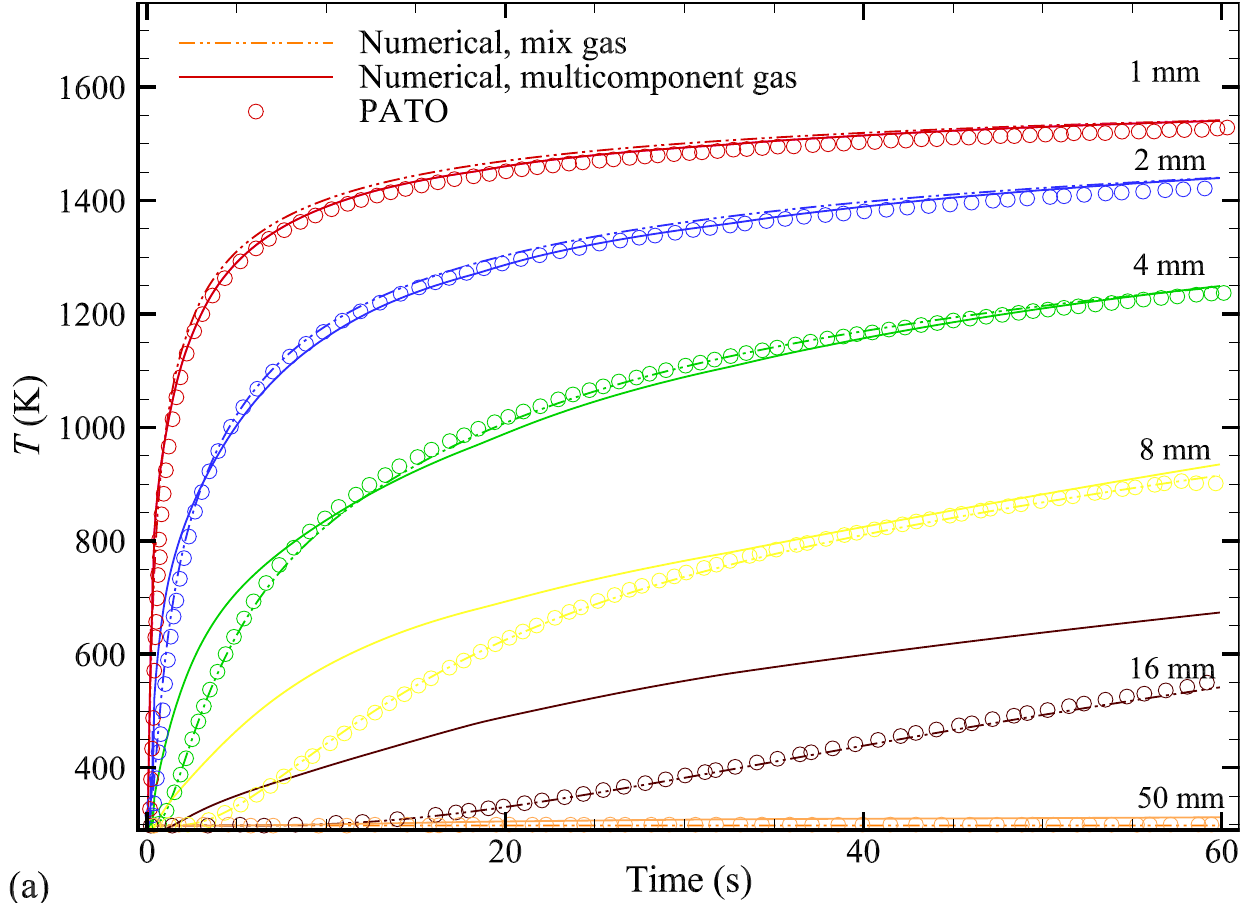}
	}	
	\subfigure
	{
		\includegraphics[width=0.45\textwidth]{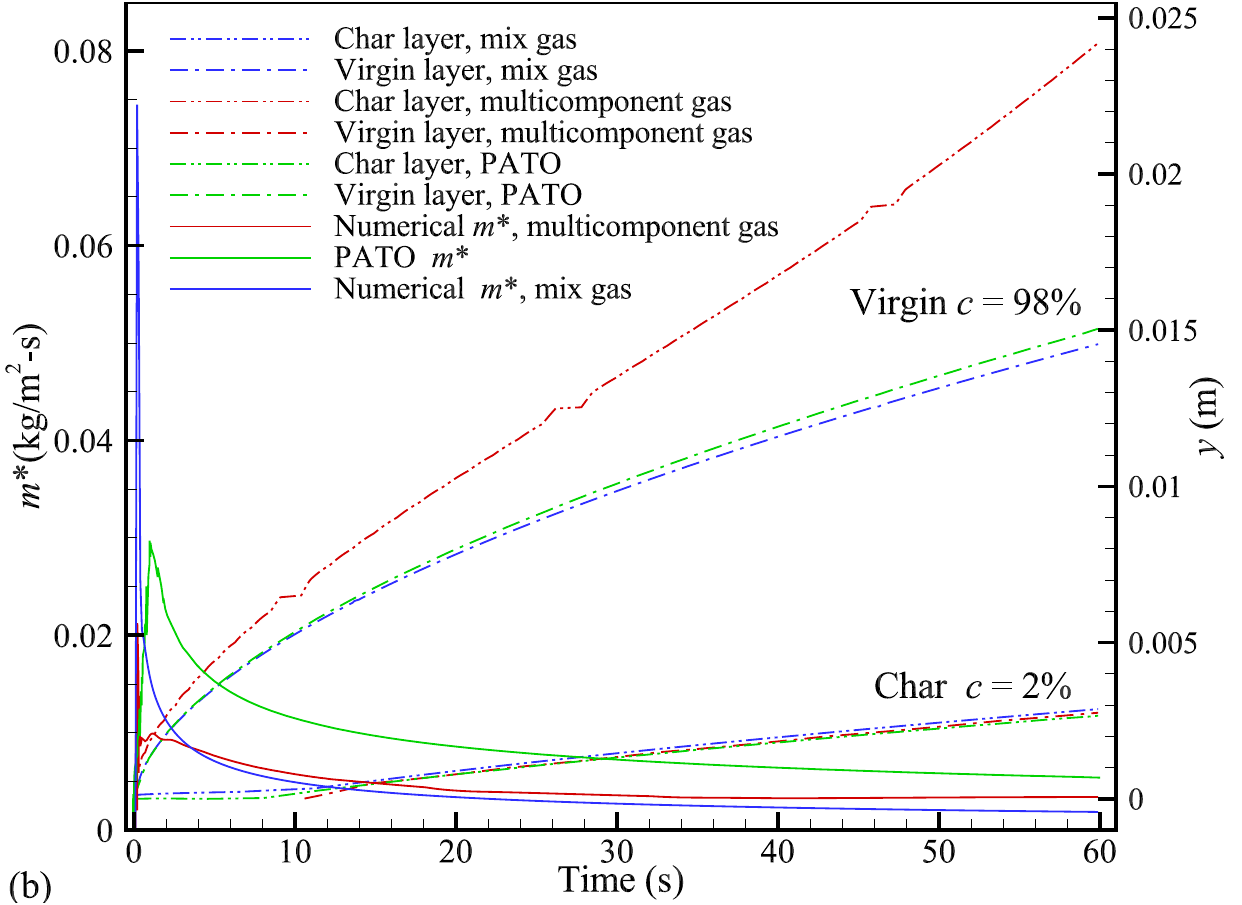}
	}	
	\caption{Comparison of the ablation quantities between the multicomponent gas model and mix gas model with the PATO results in the TACOT case:(a)temperature at different profiles over time;(b)ablation rate, char layer position, and virgin layer position over time.}
	\label{Fig.TACOTcase1-mix}
\end{figure}

\begin{figure}[htb!]
	\centering
	\subfigure
	{
		\includegraphics[width=0.45\textwidth]{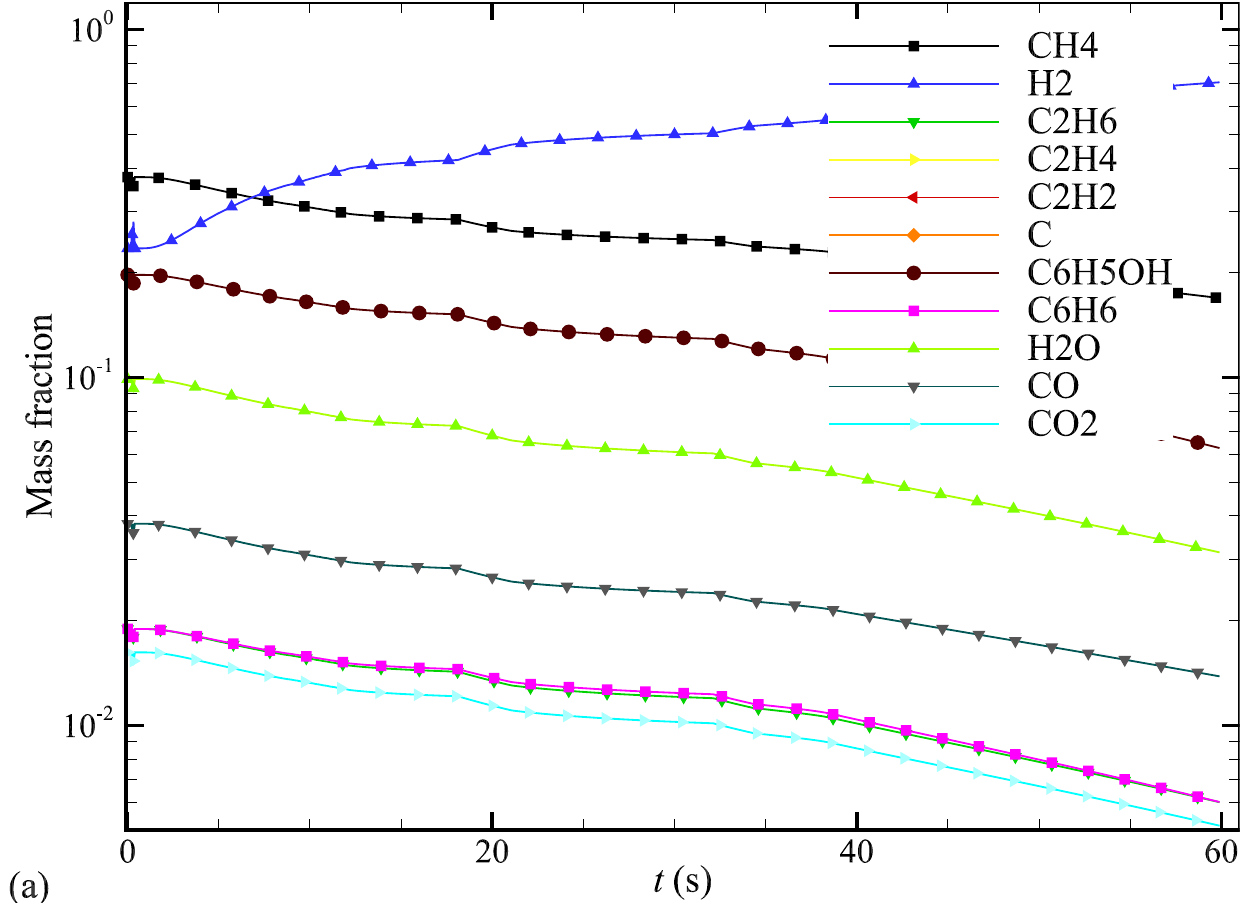}
	}	
	\subfigure
	{
		\includegraphics[width=0.45\textwidth]{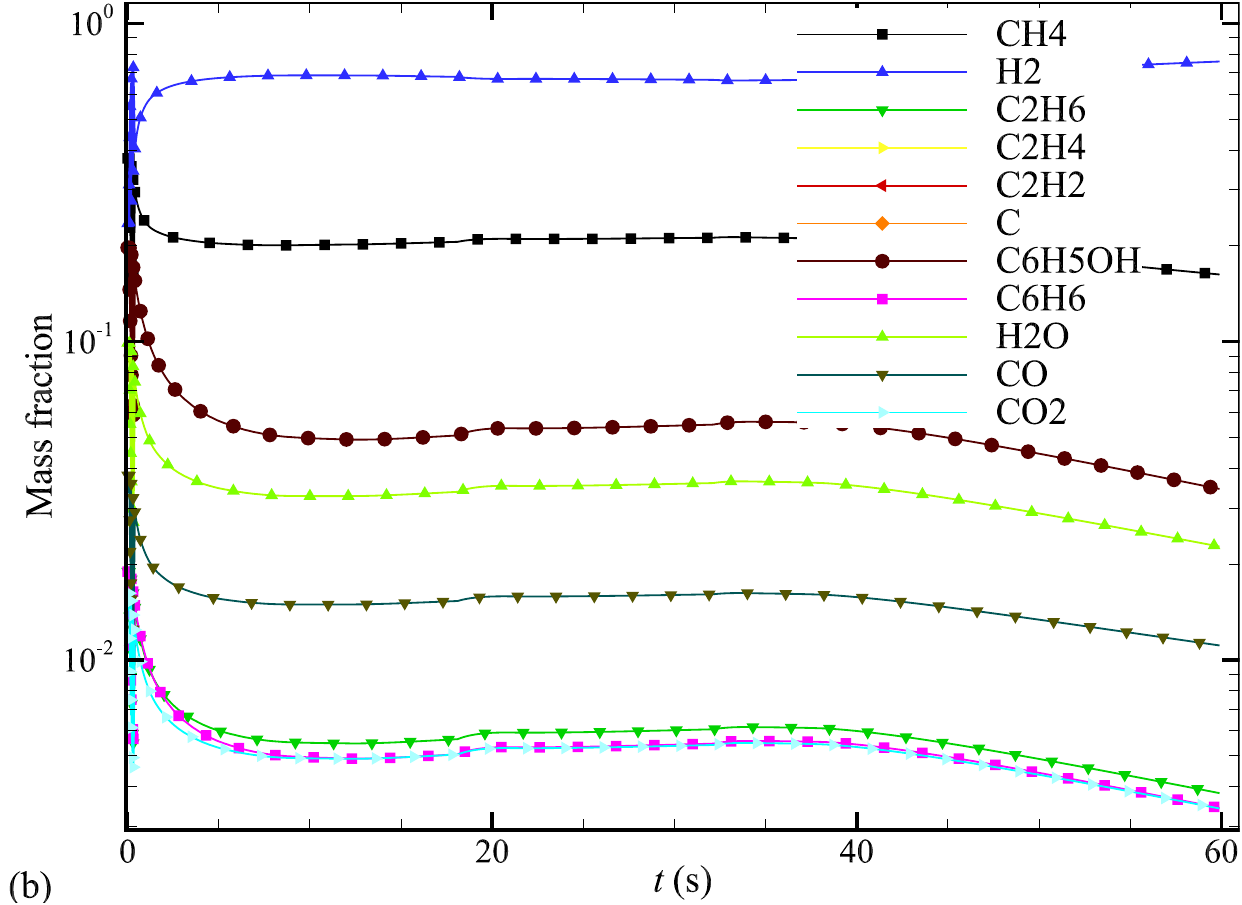}
	}	
	% \subfigure
	% {
	% 	\includegraphics[width=0.45\textwidth]{figure/TACOTcase1-tc16-y.pdf}
	% }	
	% \subfigure
	% {
	% 	\includegraphics[width=0.45\textwidth]{figure/TACOTcase1-tc50-y.pdf}
	% }	
	\caption{Mass fraction concentration of the pyrolysis gases at different profile:(a)wall;(b)4mm}
	\label{Fig.TACOTcase1-massfraction}
\end{figure}

The multicomponent gas model provides species concentration to improve the fidelity for the numerical method of the MTR. A comparison of temperature and ablation rate over time between the mix gas model and the multicomponent gas model is shown in Fig.~\ref{Fig.TACOTcase1-mix}. The multicomponent gas model results are close to the mixture gas model results near the wall, but gradually show differences in the backup region. The species concentration is close to the equilibrium concentration of the mix pyrolysis gases model near the wall where the temperature is high. Thus, the thermodynamic properties of the pyrolysis gases calculated by the multicomponent gas model are close to those calculated by the mix gas model. But the species concentration in the backup region deviates from the equilibrium concentration. Then, the temperature profile of calculated by the multicomponent gas model deviates from those calculated by the mix gas model. The depth of the char layer calculated by multicomponent gas model develops faster than those calculated by the mix gas model. The mass fraction concentration of the pyrolysis gas over time is shown in Fig.~\ref{Fig.TACOTcase1-massfraction}. The pyrolysis gases are dominated by $\mathrm{CH_4,H_2, H_2O}$. The concentration of $\mathrm{H_2}$ increases slowly near the wall and rapidly in the backup region.  

\subsection{Quasi-steady ablation on graphite blunt}
\label{sec.ablationcase}
\begin{figure}[htb!]
	\centering
		\centering
		\includegraphics[width=0.4\textwidth]{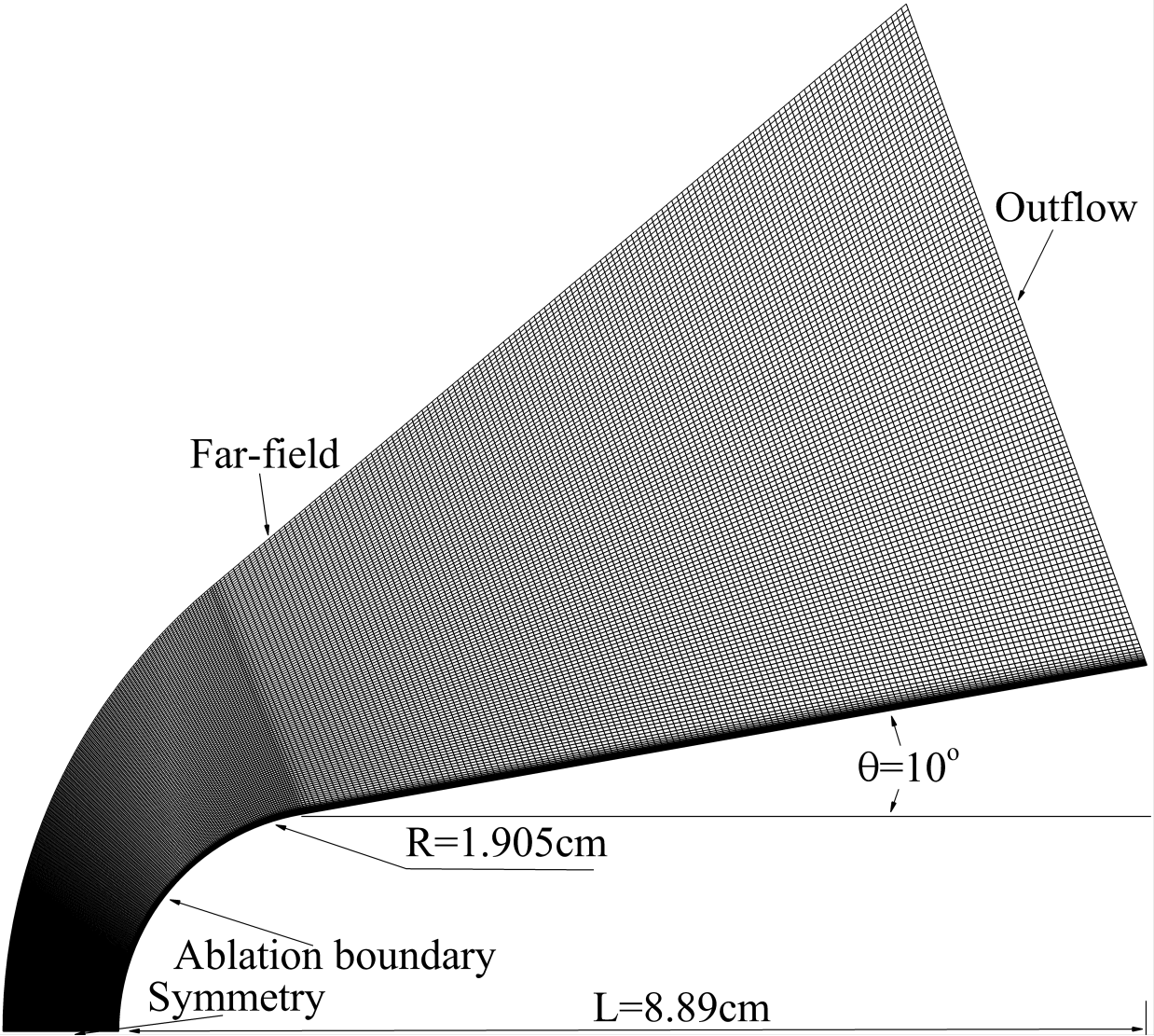}
	\caption{Illustration of the geometry and computational grid of the ARC ablation case.}
	\label{Fig.ARCgrid}
\end{figure}

To verify the accuracy of simplified ablation boundary condition, numerical tests are carried out on a graphite blunt cone with a nose radius of 1.905 cm, a cone angle of 10 $\deg$, and a total length of 8.89 cm, as shown in Fig.~\ref{Fig.ARCgrid}. This cone has been extensively tested in the Interaction Heating Facility of NASA Ames Research Center (ARC) with experimental data~\cite{doi:10.2514-1.12248} available for verification. The simplified ablation boundary condition is employed to calculate quasi-steady state ablative flow fields. The SSA assumption and RE assumption are employed in the ablation boundary condition, respectively. In the SSA assumption, $h_v = 5 \times 10^{6}$ J/kg. The surface radiation emissivity is 0.86 with a surrounding temperature of 300 K. The purpose of this case is to improve these assumptions so that the numerical results fit the experimental results better. The chemical reaction mechanism of pyrolysis gases contains 11 species and 24 reactions, while the reaction rate constants are taken from Ref.~\cite{doi:10.2514/1.J052659}. The free-stream conditions of this numerical test are taken from the experimental inflow condition, and are summarized in Table~\ref{Tab.bluntcondition}.

\begin{table}[htb!]
	\centering
	\caption{Free-stream conditions for numerical simulations of graphite blunt.}
	\label{Tab.bluntcondition}%添加标题 设置标签
	\begin{tabular}{cccccccc}
		\toprule
		%\hline
		$u$(m/s)&$\rho \mathrm{(kg/m^3)}$&$T(\mathrm{K})$&$Y_{\mathrm{N2}}$&$Y_{\mathrm{O2}}$&$Y_{\mathrm{NO}}$&$Y_{\mathrm{N}}$&$Y_{\mathrm{O}}$\\
		\midrule
		5354&$3 \times 10^{-3}$&1428&0.6169&0.0&0.0046&0.1212&0.2573\\
		%\hline
		\bottomrule
	\end{tabular}
\end{table}

\begin{figure}[htb!]
	\centering
	\subfigure
	{
		\includegraphics[width=0.46\textwidth]{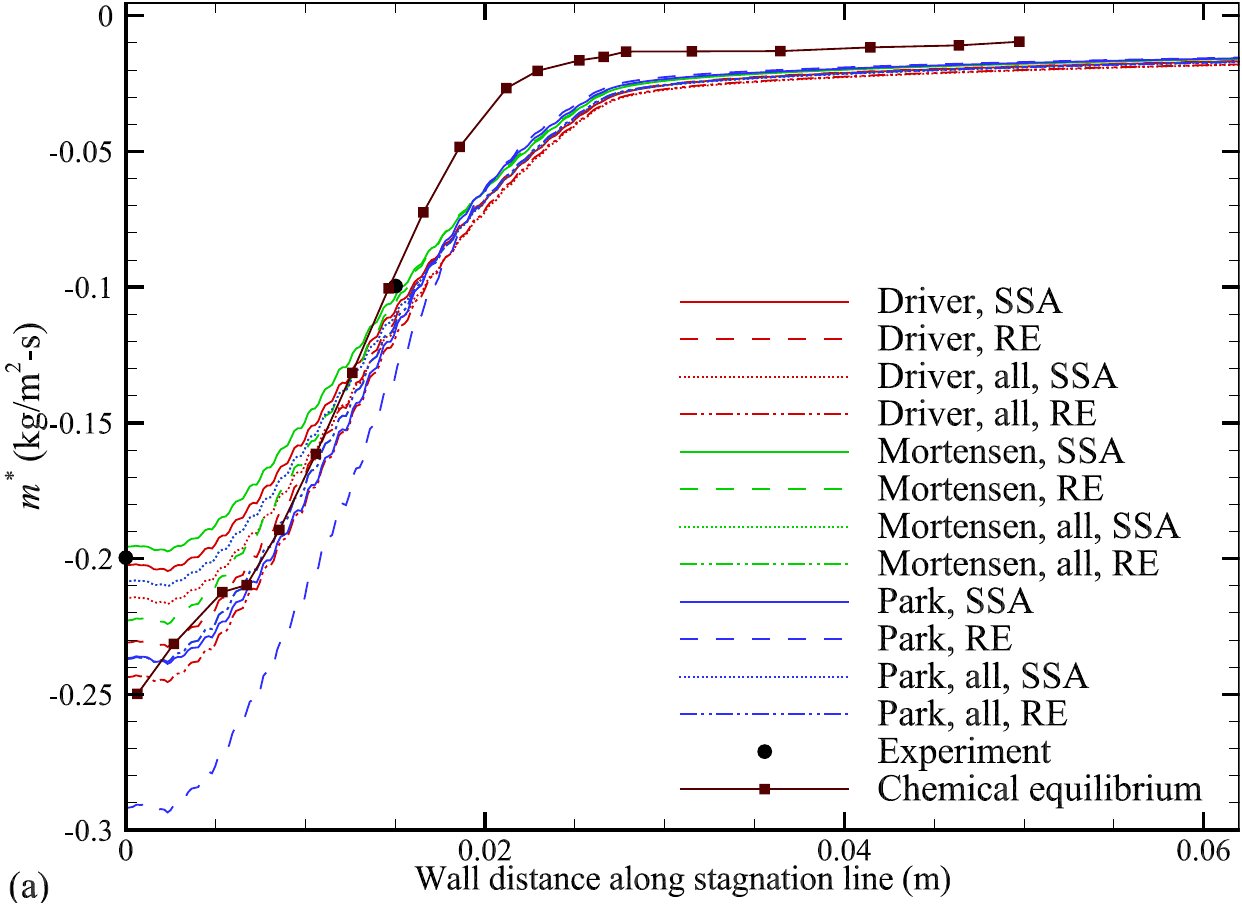}
	}	
	\subfigure
	{
		\includegraphics[width=0.46\textwidth]{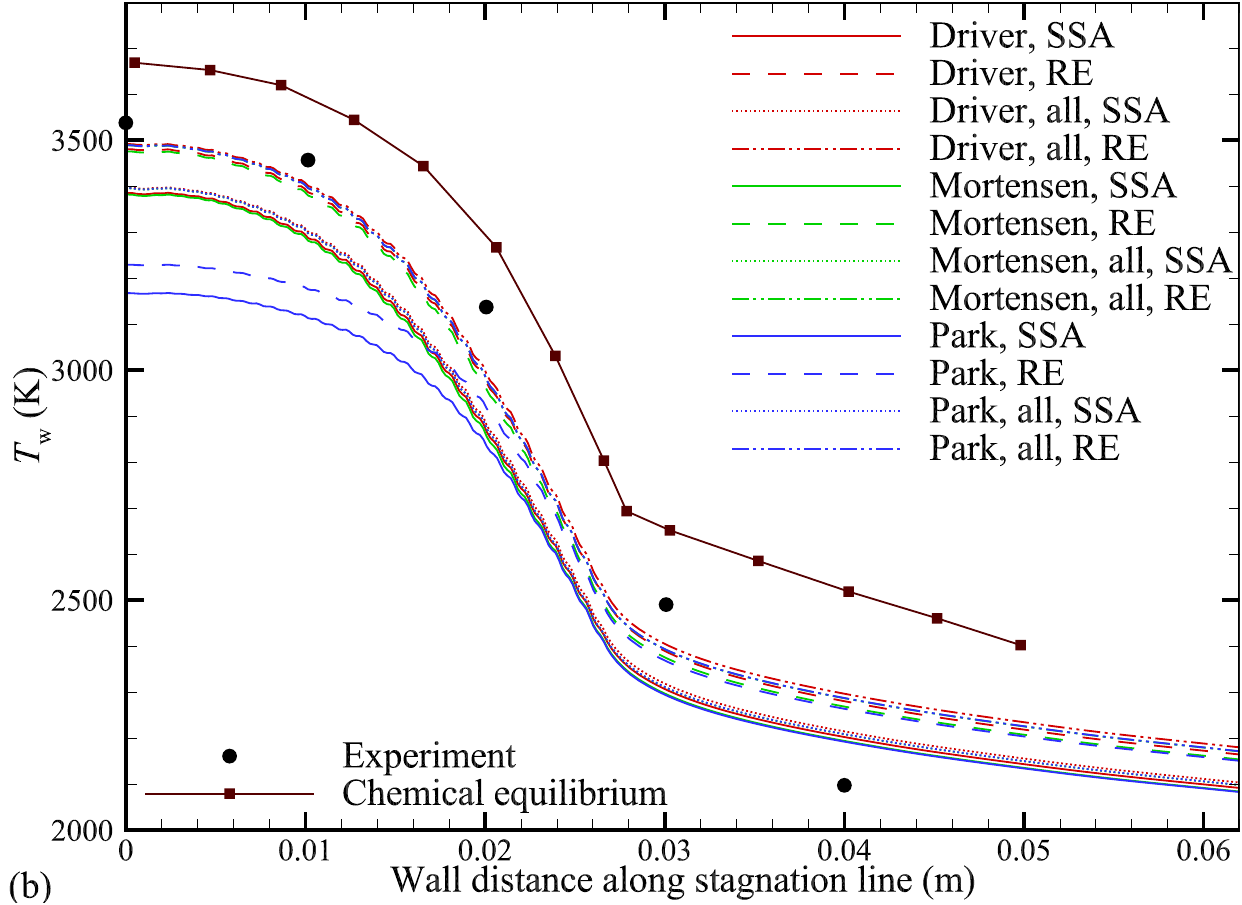}
	}	
	\caption{Comparison of the ablation quantities in the numerical simulation of the ARC graphite blunt using the simplified ablation condition, where all denotes the calculation that takes into account the nitridation and catalysis in the surface chemistry, otherwise it indicates it is not taken into account:(a)surface ablation rate and (b)surface temperature.}% Mortensen mechanism use the same surface chemistry with Park except sublimation.} or the opposite if all is not included
	\label{Fig.graphite}
\end{figure}

\begin{figure}[htb!]
	\centering
	\subfigure
	{
		\includegraphics[width=0.46\textwidth]{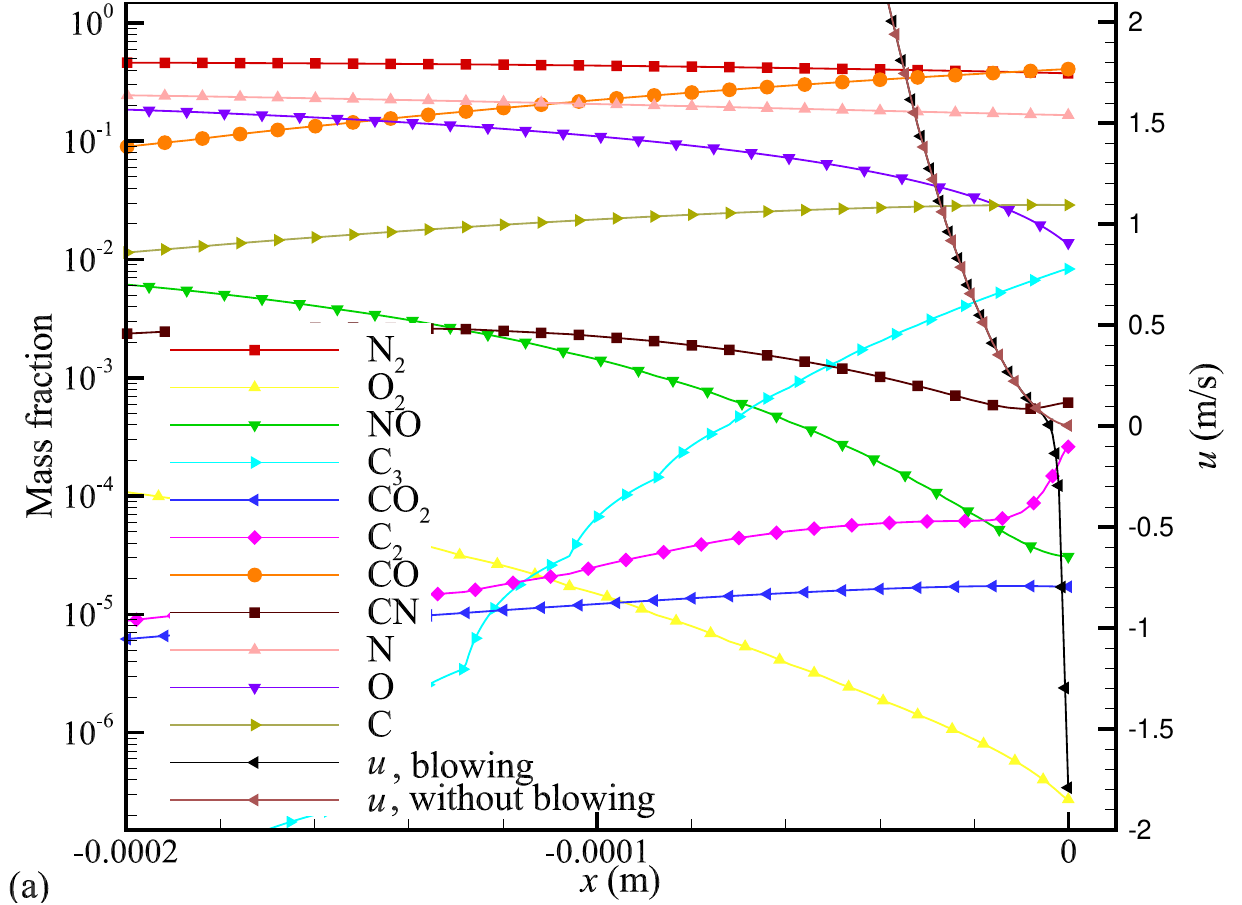}
	}	
	\subfigure
	{
		\includegraphics[width=0.46\textwidth]{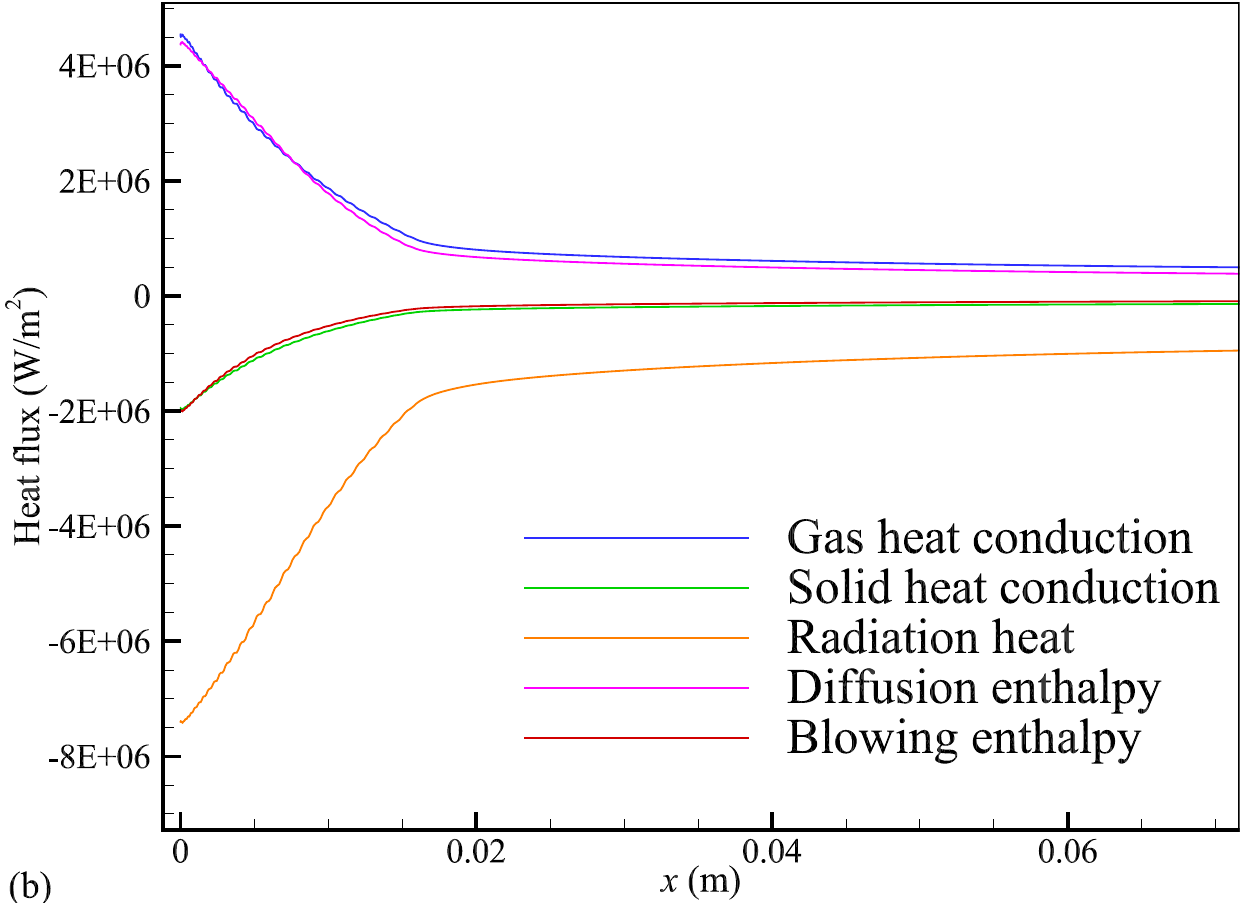}
	}	
		\caption{Numerical results of the ARC graphite blunt:(a)specie concentration profile and velocity profile along the central axis;(b)distribution of heat in SEB along the surface.}
	\label{Fig.graphiteprofile}
\end{figure}

A comparison between the numerical results and experimental data is show in Fig.~\ref{Fig.graphite}, including results from different surface chemistry models and chemical equilibrium model~\cite{doi:10.2514-1.47995}. Notably, the results of the Mortensen model without the nitridation and catalysis show good agreement with the experimental data in both ablation rate and surface temperature. If the nitridation and catalysis are considered, the ablation rate is higher than the experimental data indicating that the nitridation rate is overestimated. If the RE assumption is used, the calculated results of all surface chemistry mechanisms show lower temperature and higher ablation rates compared to the results using the SSA model. The SSA assumption provides a reasonable approximation to the MTR that improves the accuracy of the results and fits the experimental data better. Chemical equilibrium result shows higher ablation rate because the equilibrium assumption overestimates the reaction rate and the actual reaction rate is limited by the diffusion of oxygen. The simplified ablation boundary using the SSA assumption and the Mortensen reaction mechanism can accurately calculates the ablation rate, which is the closest to the experimental data. In addition, Fig.~\ref{Fig.graphiteprofile} compares the specie concentration and velocity profiles on the wall. In the case of considering the blowing, ablation production gases are injected to the flow fields with a blowing velocity of about 2 m/s. The velocity profiles are consist whether or not the blowing is considered except near the wall. The species concentration of the ablation product (e.g., $\rm{CO}$, $\rm{C_3}$) is high on the wall and decreases with increasing height from wall. The component distribution of the surface energy balance on the wall shows that most of the heat load is dissipated by radiation. The convection enthalpy of surface blowing also helps to offset the heat load.   
%The oxidation,catalyst, and sublimation are considered in th surface chemistry to evaluate the ablation production rate. a velocity of 5354 m/s, a density of 0.003$\rm{kg/m^3}$, a temperature of 1428 K, and a mass fraction of $\rm{O_2, N_2, NO, N, O}$ being 0.0, 0.6169, 0.0046, 0.1212, 0.2573.


\subsection{Strong coupled simulation between CRF and MTR}
\begin{figure}[htb!]
	\centering
		\centering
		\includegraphics[width=0.5\textwidth]{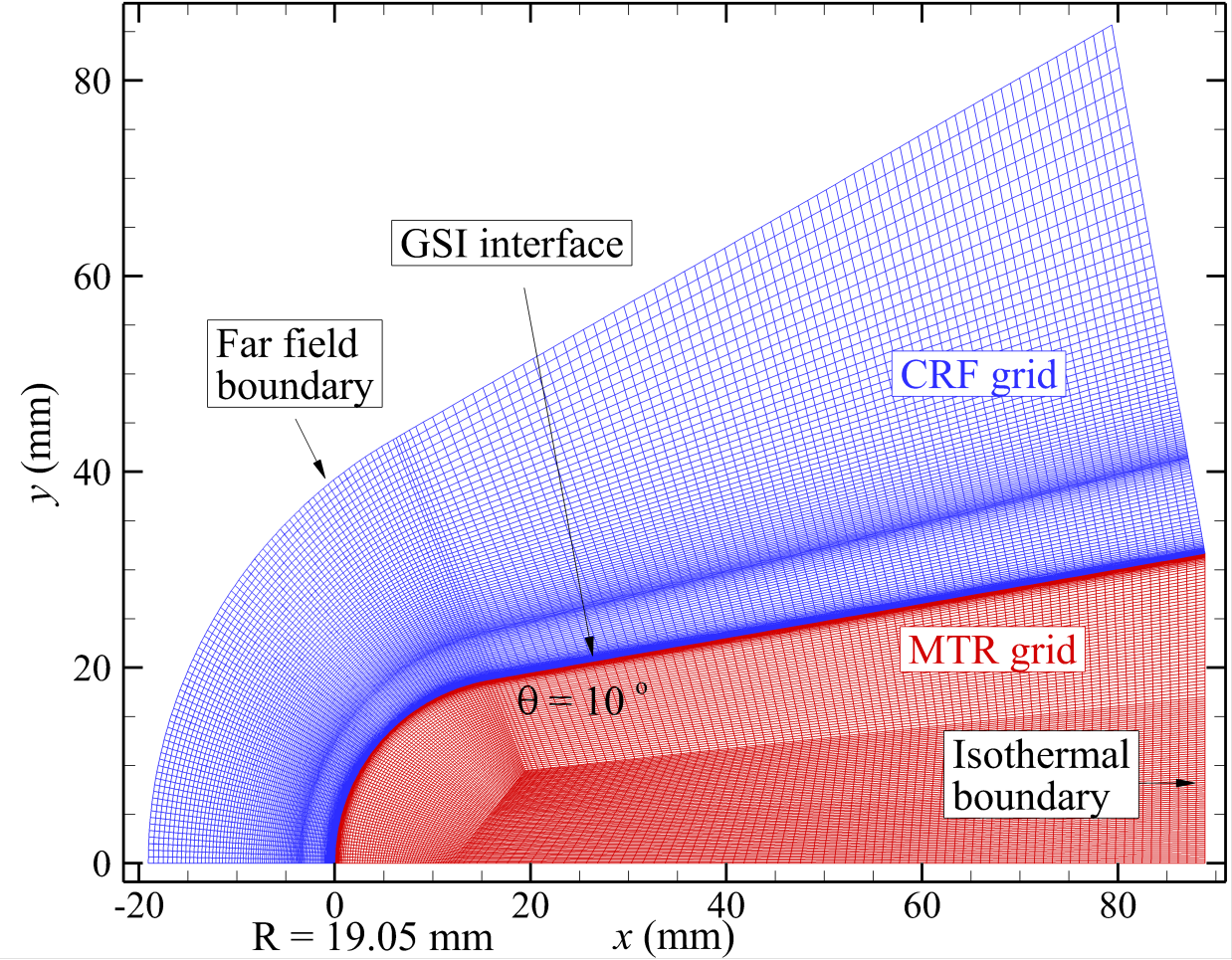}
	\caption{The schematic of the computational grid and boundary of the ARC blunt.}
	\label{Fig.ARCcoupledgrid}
\end{figure}

% 这个算例目的在于验证隐式耦合方法能够提高非定常计算精度，对比接偶计算和
Coupled simulations between the CRF and the MTR are conducted on the ARC blunt. The inflow conditions are the same as the experimental conditions, as given in Table~\ref{Tab.bluntcondition}. The computational grid contains a fluid part and a material part, as shown in Fig.\ref{Fig.ARCcoupledgrid}. Different to the quasi-steady ablation case in Section~\ref{sec.ablationcase}, the strong coupled simulation solves $q_{cond}$ by the MTR solver here to calculate the variant of wall temperature. A physical time step of 0.1 s is employed both for CRF and MTR. The material properties are taken from the PICA material parameters, as given in Ref.\cite{TACOTv3.0}. The explicit and implicit coupling mechanisms, as illustrated in the Fig.\ref{Fig.coupling} are employed and compared based on their respective results. 

\begin{figure}[htb!]
	\centering
	\subfigure
	{
		\centering
		\includegraphics[width=0.45\textwidth]{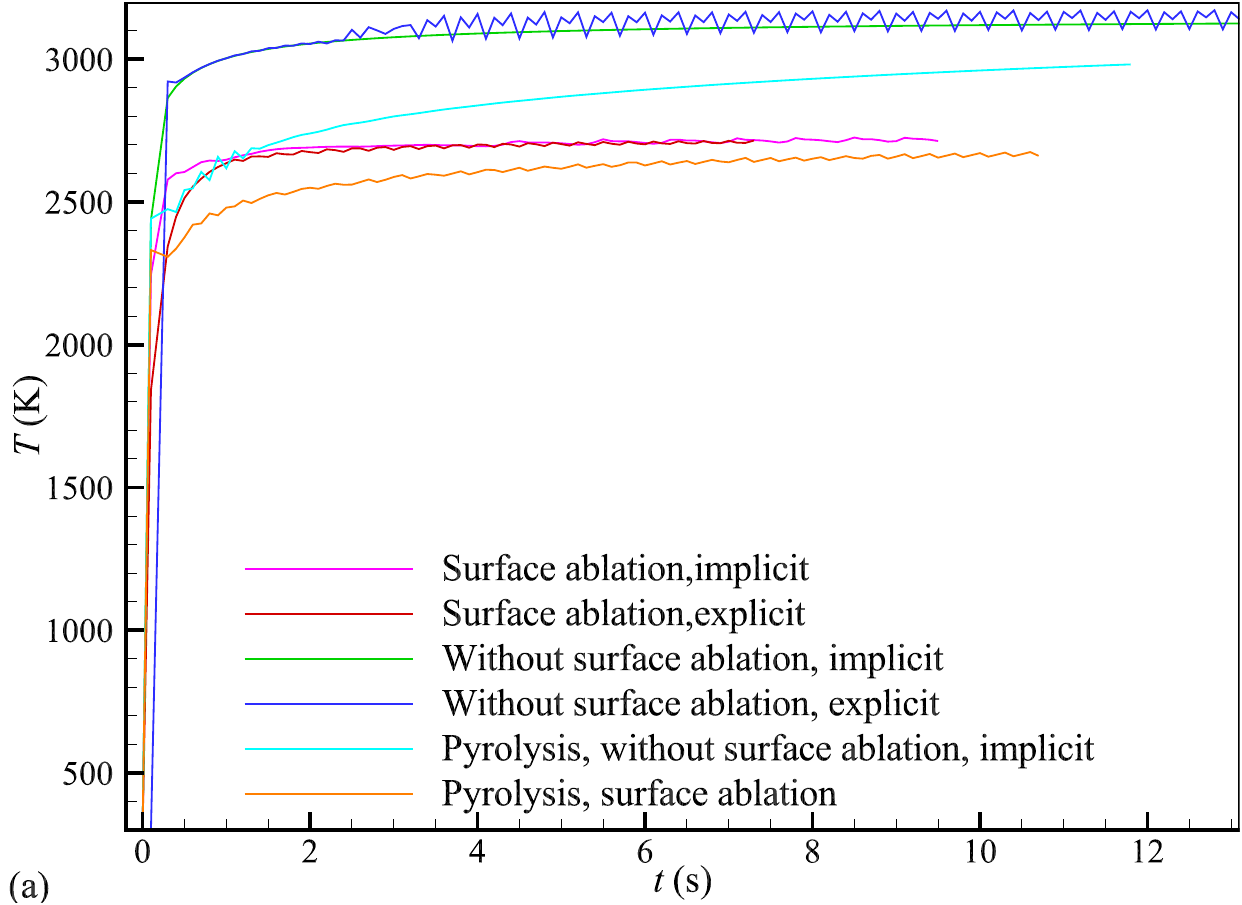}
	}	
	\subfigure
	{
		\centering
		\includegraphics[width=0.45\textwidth]{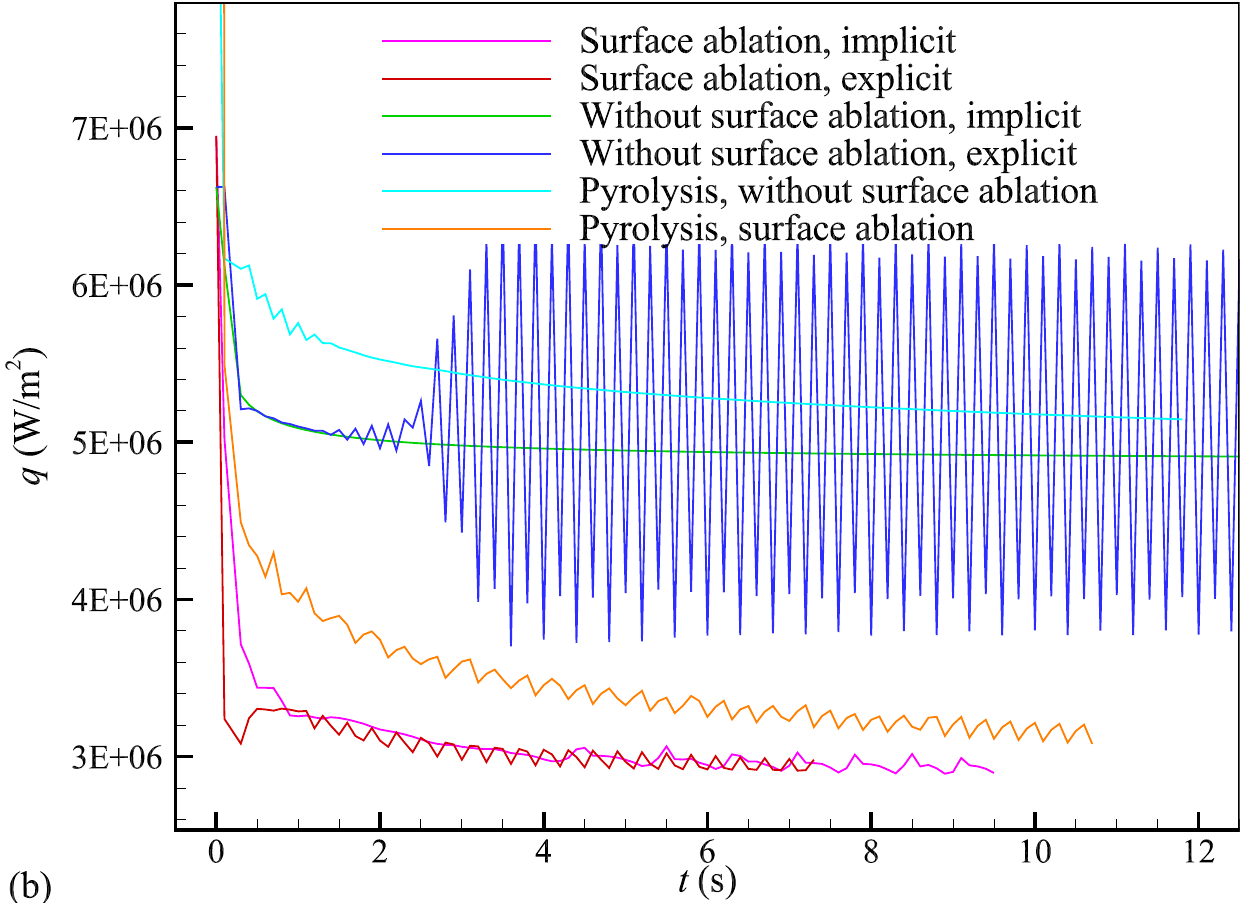}
	}	
	\caption{Comparison of surface quantities at the nose of the ARC blunt in the coupled simulation between the CRF and MTR:(a)surface temperature;(b)heat flux.}
	\label{Fig.ARC-compare}
\end{figure}

\begin{figure}[htb!]
	\centering
	\subfigure
	{
		\centering
		\includegraphics[width=0.45\textwidth]{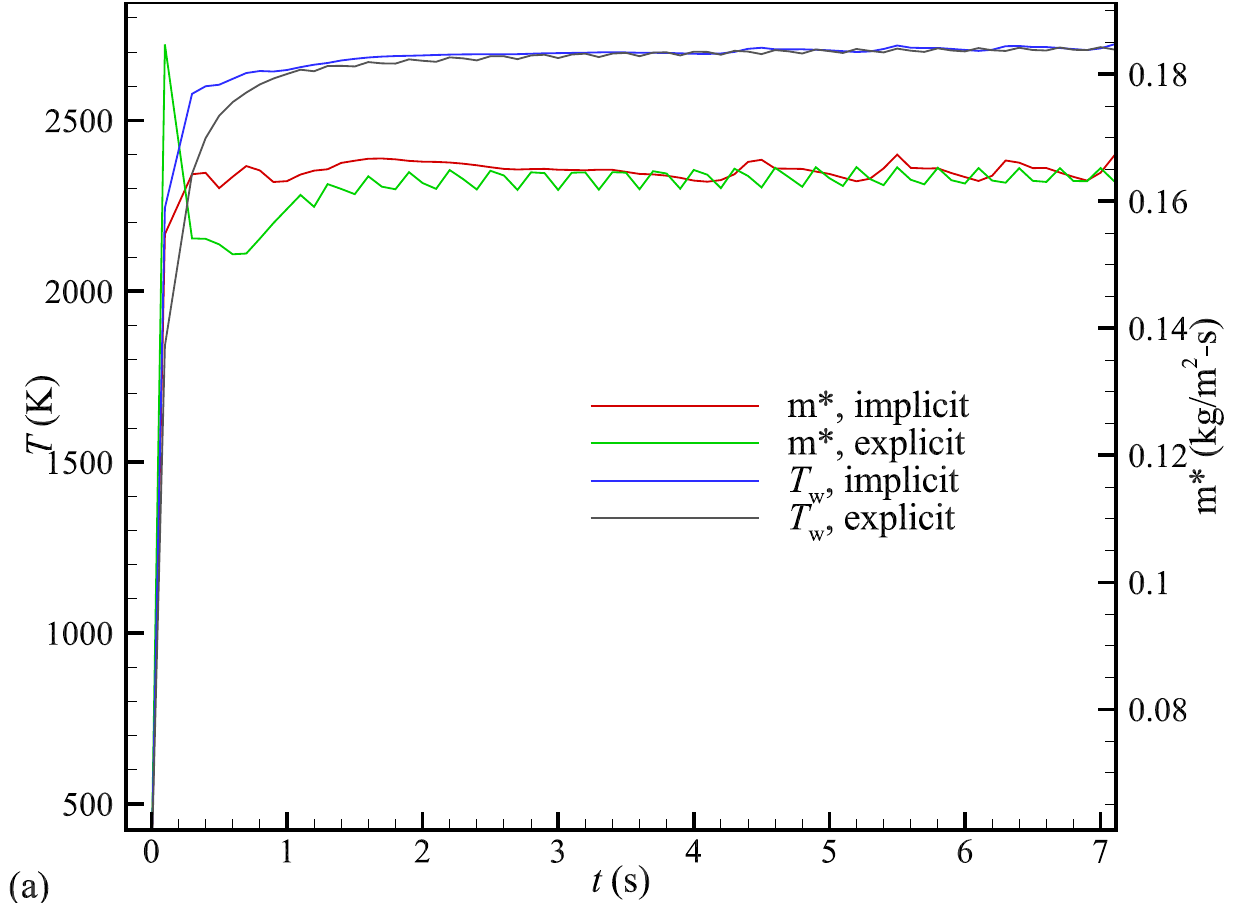}
	}	
	\subfigure
	{
		\centering
		\includegraphics[width=0.45\textwidth]{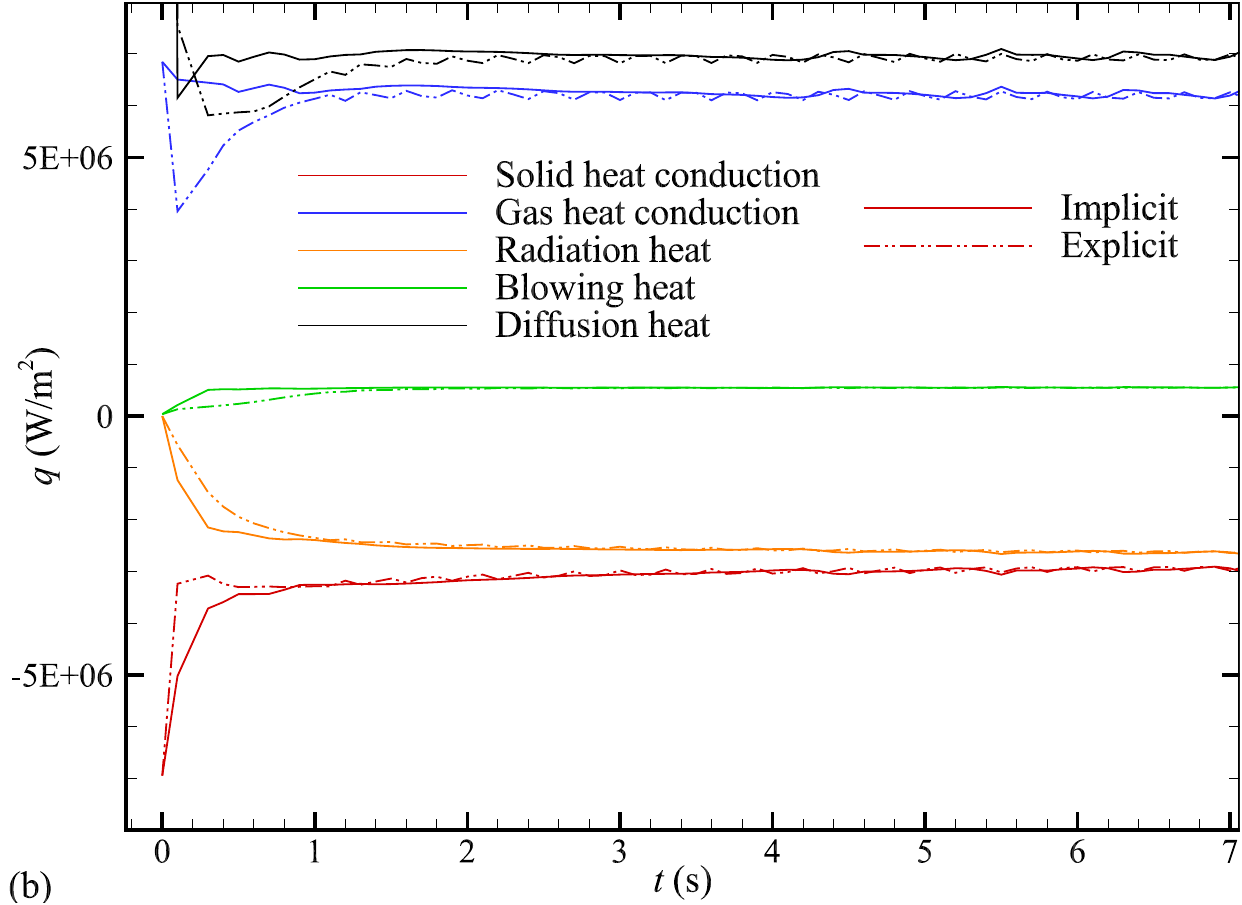}
	}	
	\caption{Comparison of ablation parameters at the nose of the ARC blunt in the coupled simulation between the CRF and MTR:(a)surface temperature and ablation rate;(b)energy distribution on the surface.}
	\label{Fig.ARC-compare-ablation}
\end{figure}

\begin{figure}[htb!]
	\centering
	\subfigure
	{
		\centering
		\includegraphics[width=0.45\textwidth]{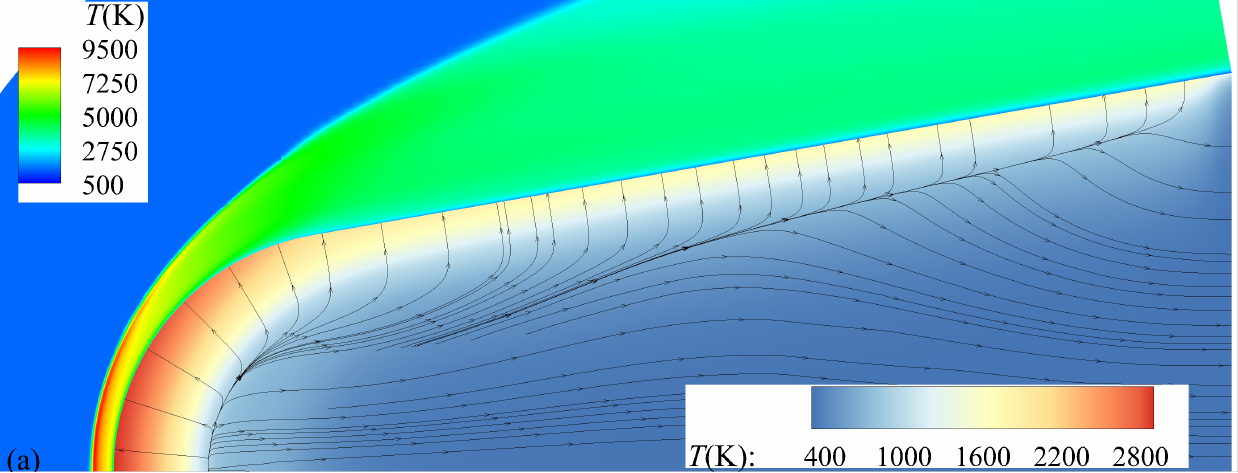}
	}	
	\subfigure{
		\includegraphics[width=0.45\textwidth]{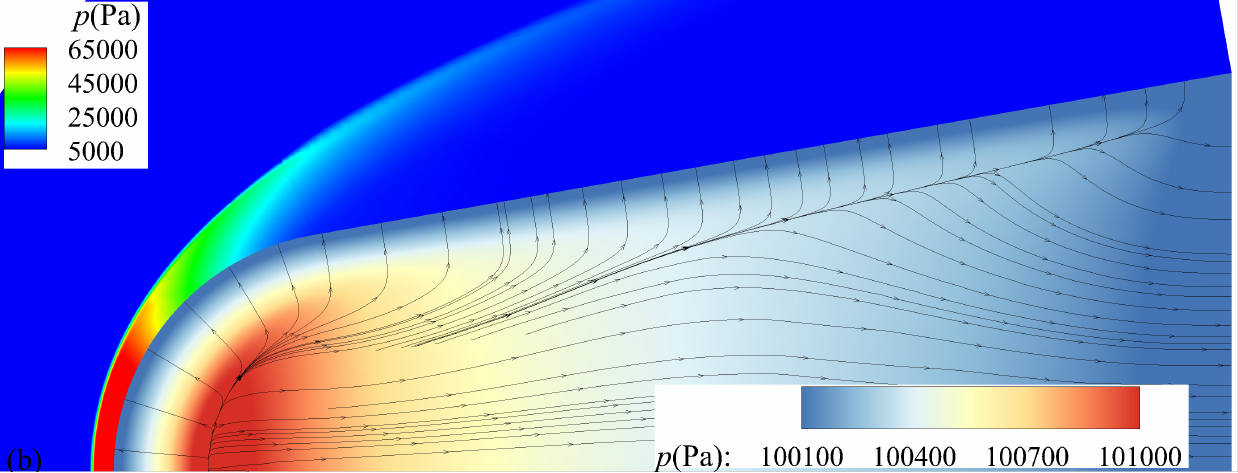}
	}	
	\subfigure{
		\centering
		\includegraphics[width=0.45\textwidth]{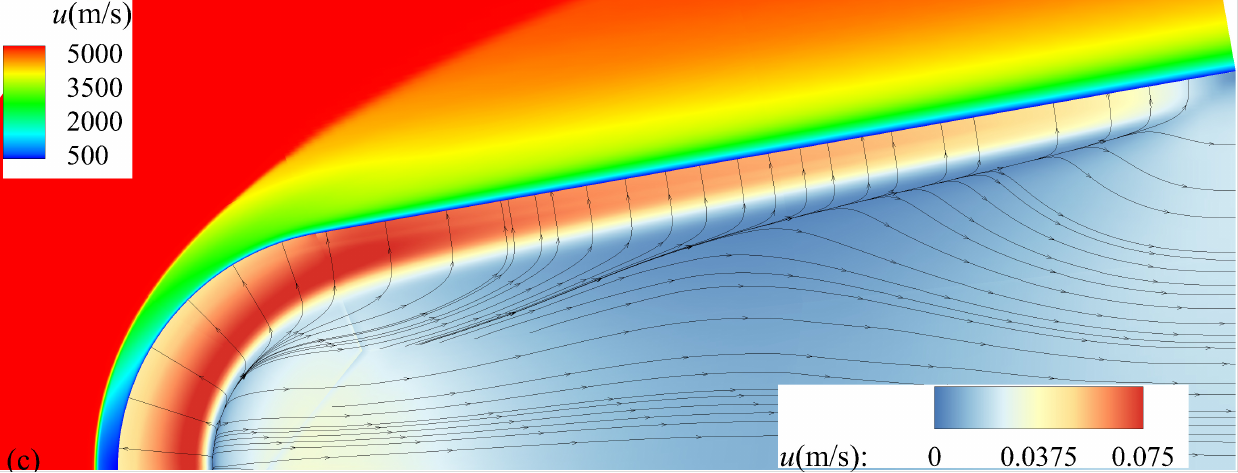}
	}	
	\subfigure{
		\centering
		\includegraphics[width=0.45\textwidth]{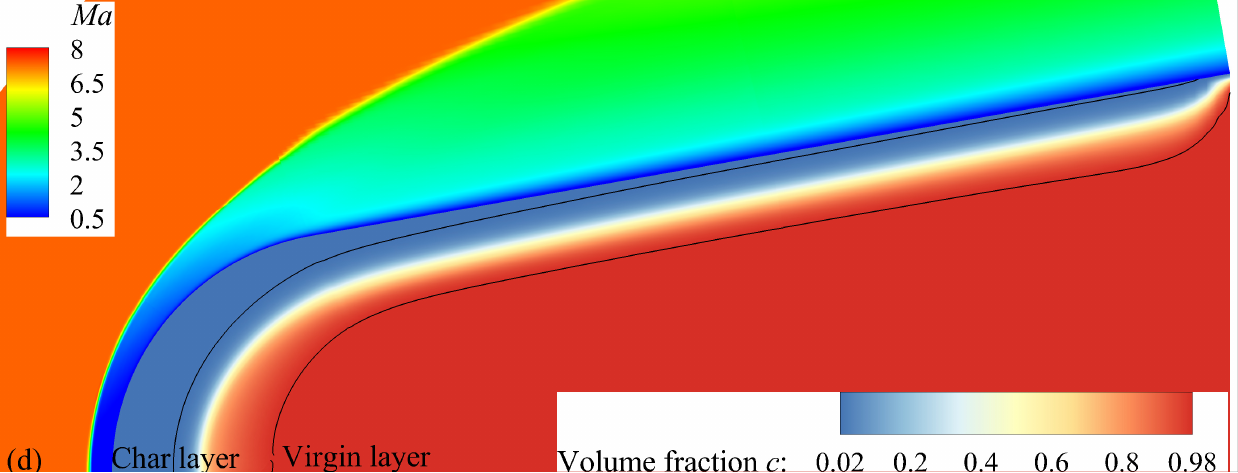}
	}	
	% {
	% 	\includegraphics[height=0.1\textwidth]{figure/ARC-ablation-t16.pdf}
	% }	
	% \includegraphics[width=0.45\textwidth]{manu/ARC-Tw.pdf}
	\caption{Contour of flow fields and material parameters at $t=16$ s in the coupled simulation between the CRF and MTR:(a) temperature;(b) pressure;(c) velocity magnitude;(d) Mach number of flow fields and volume fraction of charring materials.}
	\label{Fig.ARC-pyrolysis-contour}
\end{figure}

The wall temperature and heat flux of the explicit coupling mechanism and the implicit coupling mechanism are compared in Fig.~\ref{Fig.ARC-compare}. The results of the explicit coupling mechanism show significant oscillations in wall temperature, while the implicit coupling method effectively suppresses the oscillations because the physical quantities on the wall are exchanged at the pseudo time level and are continuously changed. The wall temperature of the explicit and implicit coupling mechanisms are roughly the same, but the explicit coupling method is slightly lag behind the implicit coupling method. When the surface ablation is considered, the surface temperature at the same time step is lower because the surface reactions carries away heat flux through mass loss. This conclusion can be verified in Fig.~\ref{Fig.ARC-compare-ablation} that the surface ablation introduces blowing heat to offset the heat load. In addition, when the pyrolysis gases is taken into account, the surface temperature at the same time step is lower than the results without considering pyrolysis. This is because the pyrolysis gases carry away the heat load and reduce the thermal conductivity. When the pyrolysis and ablation is considered, the contour of the flow fields and material parameters are shown in Fig.~\ref{Fig.ARC-pyrolysis-contour}. The velocity magnitude of the pyrolysis gases within the ablative material is low, with a maximum value of 0.075 m/s. When the material temperature increase, the resin matrix in the charring materials gradually pyrolyzes into gases, reducing the volume fraction of the ablative material. Meanwhile, the thickness of the virgin layer gradually increases. And the virgin layer turns into the char layer eventually.

\subsection{Coupled simulation of reentry capsule}

\begin{figure}[htb!]
	\centering
		\includegraphics[width=0.5\textwidth]{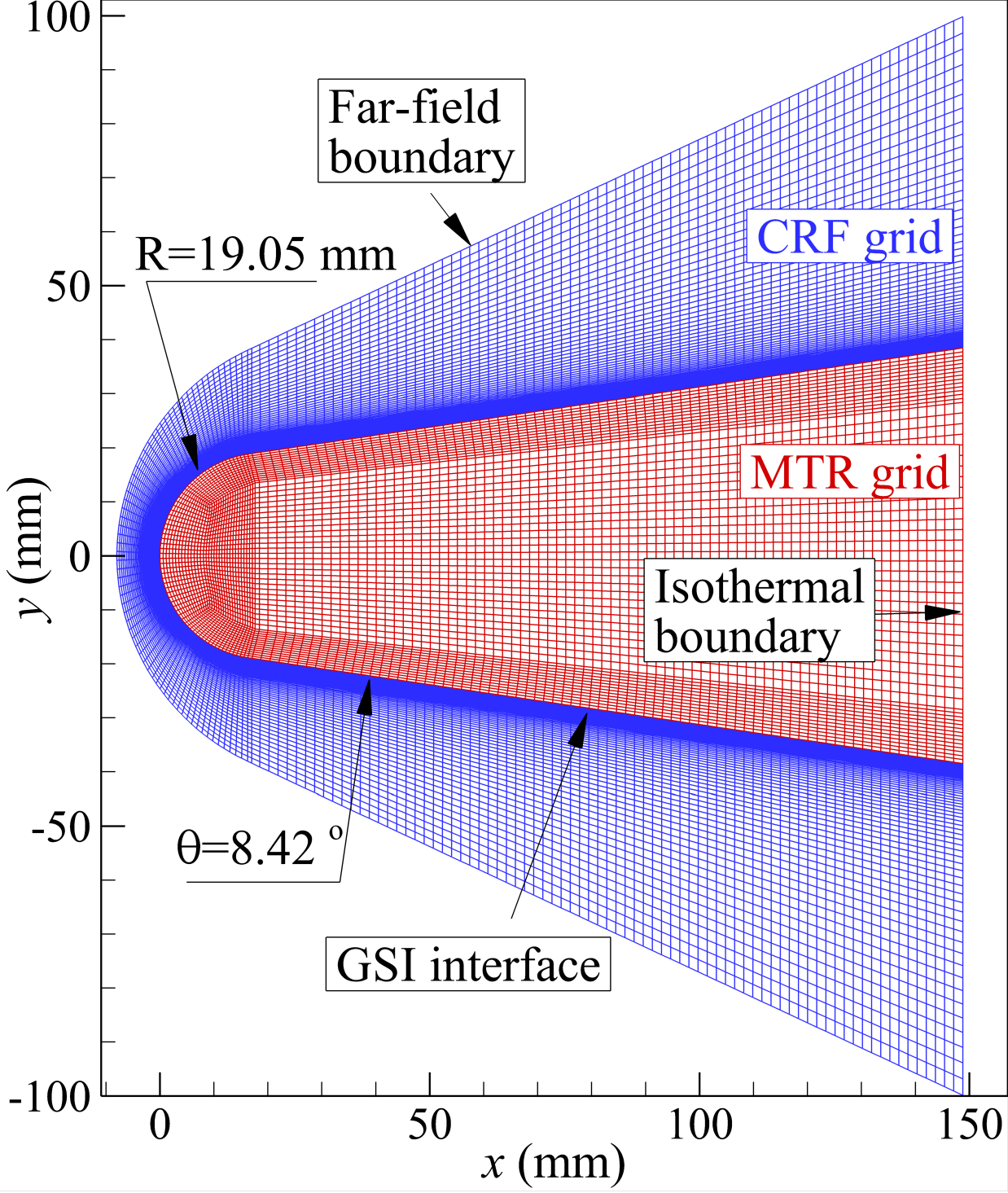}
	\caption{Illustration of geometry, grid, and boundary setting of the IRV case.}
	\label{Fig.IRV-illustration}
\end{figure}

The explicit and implicit coupling mechanism are used to simulate the unsteady heating process of the thermal protection system material of the IRV-2 vehicle during its flight trajectory. The IRV-2 is an axisymmetric sphere-biconic reentry with a nose radius of 1.905 cm, a biconic angle of 8.42$^{\circ}$, and a length of 0.1488 m. The computational domain and numerical setup are shown in Fig.~\ref{Fig.IRV-illustration}. The unsteady solutions are solved by the dual time-step method with a physical time step of 0.25 s. The free-stream conditions at different time steps are derived by linear extrapolation of the flight conditions of the trajectory, which is given in Table~\ref{Tab.trajectory}. The TPS material of the IRV-2 reentry is $\rm{ZrB_2}$~\cite{doi:10.1061} which does not undergo pyrolysis. The material properties of $\rm{ZrB_2}$ are given as follows: a density of 6000 $\rm{kg/m^3}$, a specific heat of 628 $\rm{J/(kg \cdot K)}$, and a thermal conductivity of 66 $\rm{W/(m \cdot K)}$. The surface radiation emissivity is 0.8 with a surrounding temperature of 300 K. The GSI interface is used to simulate the variation of the interfacial quantities between the CRF and MTR solutions, while the catalysis is not considered in the surface chemistry. The explicit method introduces a time delay error and gradually deviates from the experimental data with increasing time. Thus, time step is constrained to reduce the time discretization error, resulting in a significant increase in computational time cost. However, the implicit coupling method eliminates the time delay error by exchanging interfacial quantities in the pseudo-iteration step, allows a larger time step while maintaining high accuracy.

\begin{figure}[htb!]
	\centering
	\subfigure
	{
		\includegraphics[width=0.45\textwidth]{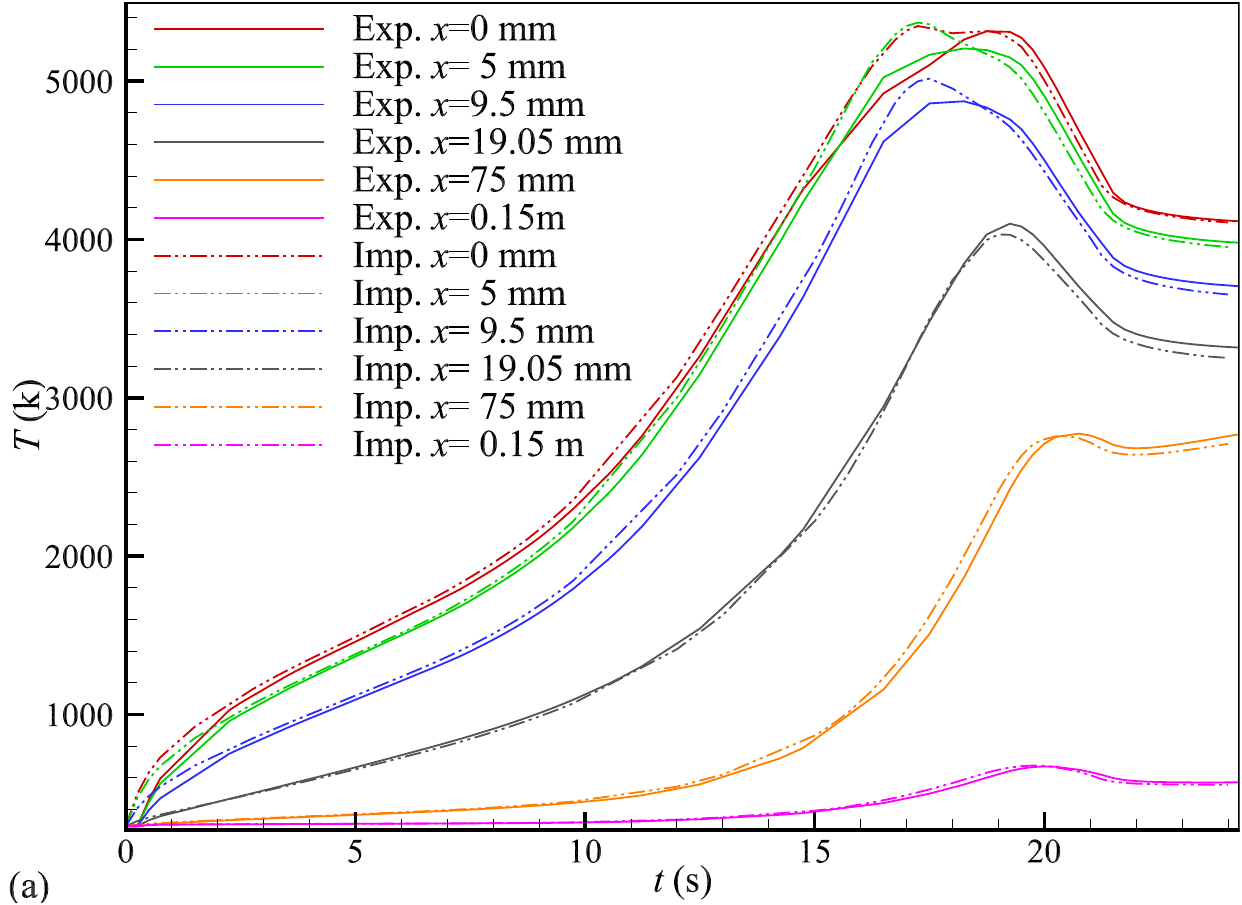}
	}	
	\subfigure
	{
		\includegraphics[width=0.45\textwidth]{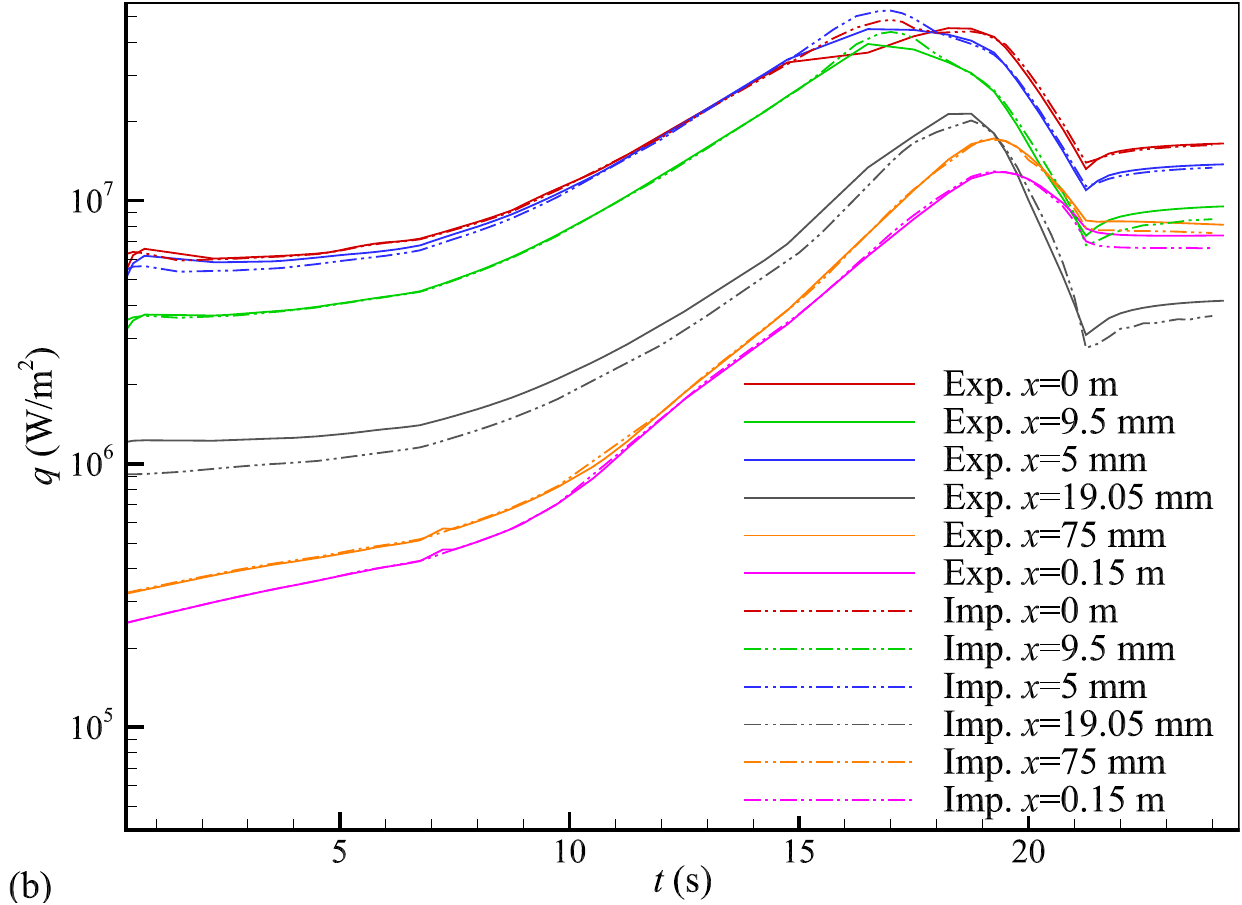}
	}
		% \centering
		% \includegraphics[width=0.6\textwidth]{figure/coupling.pdf}
	\caption{Comparison of the surface quantities between the explicit coupling and implicit coupling mechanisms in the coupled CRF and MTR simulation of the IRV-2 reentry:(a) surface temperature; (b) surface heat flux.}
	\label{Fig.IRV2coupling}
\end{figure}

\begin{figure}[htb!]
	\centering
	\subfigure
	{
		\includegraphics[width=0.45\textwidth]{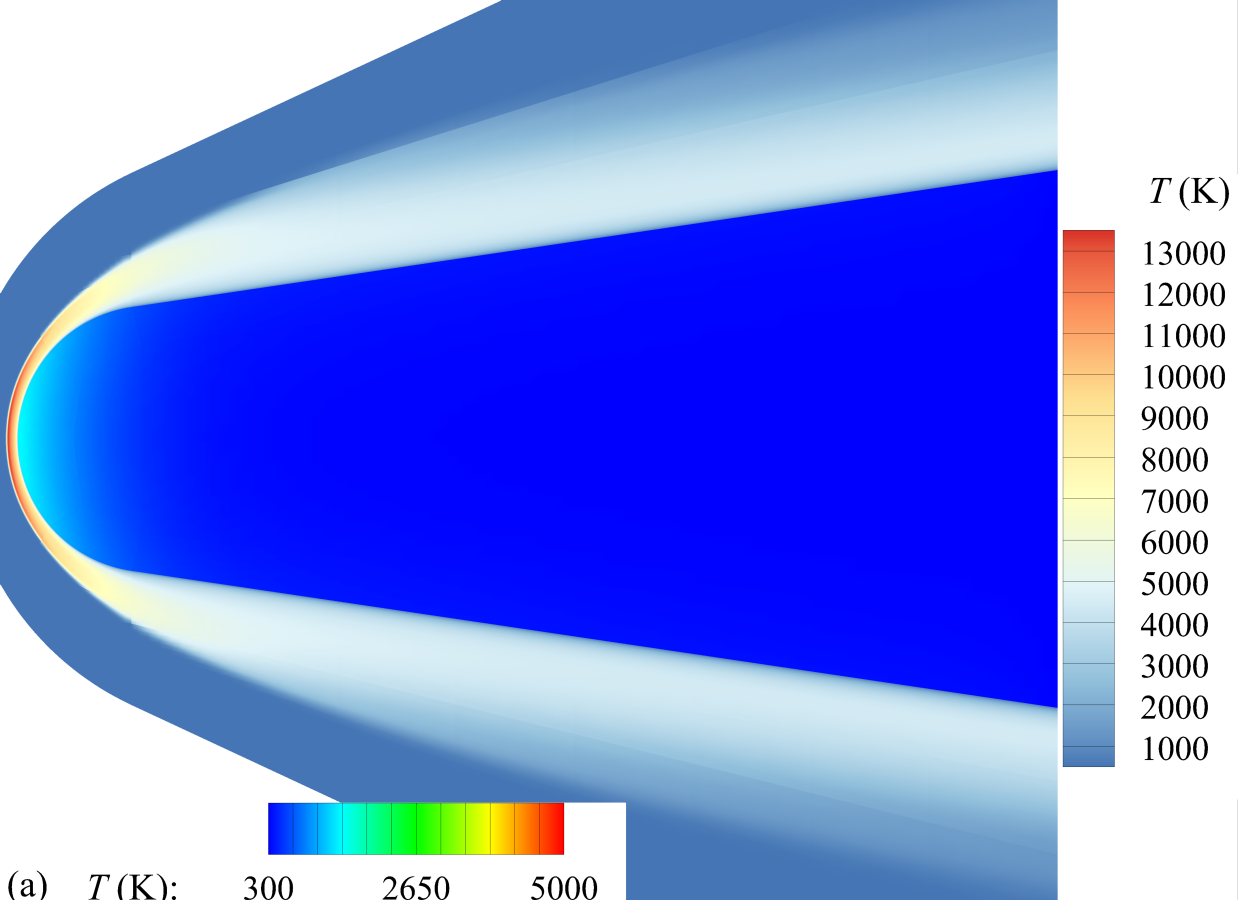}
	}	
	\subfigure
	{
		\includegraphics[width=0.45\textwidth]{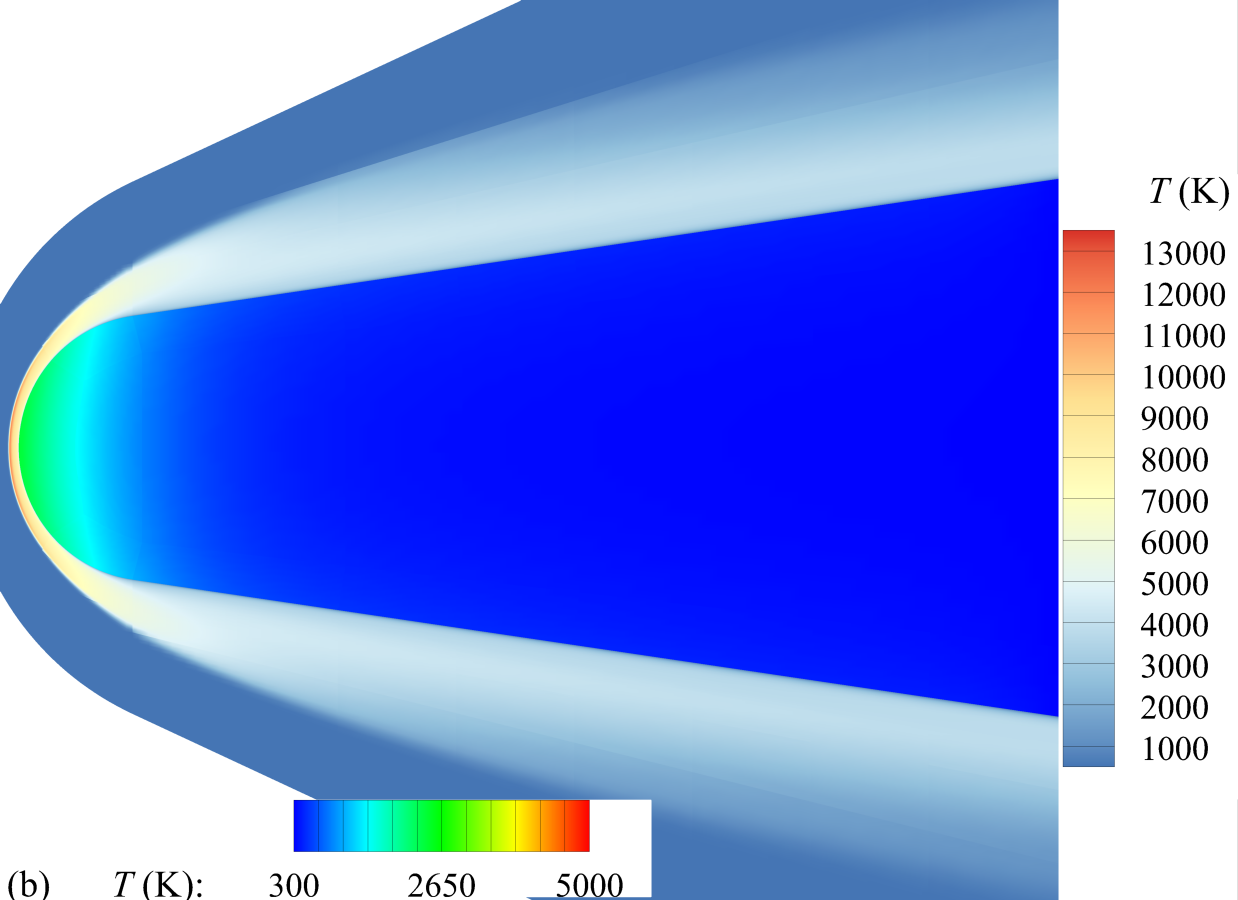}
	}
	\subfigure
	{
		\includegraphics[width=0.45\textwidth]{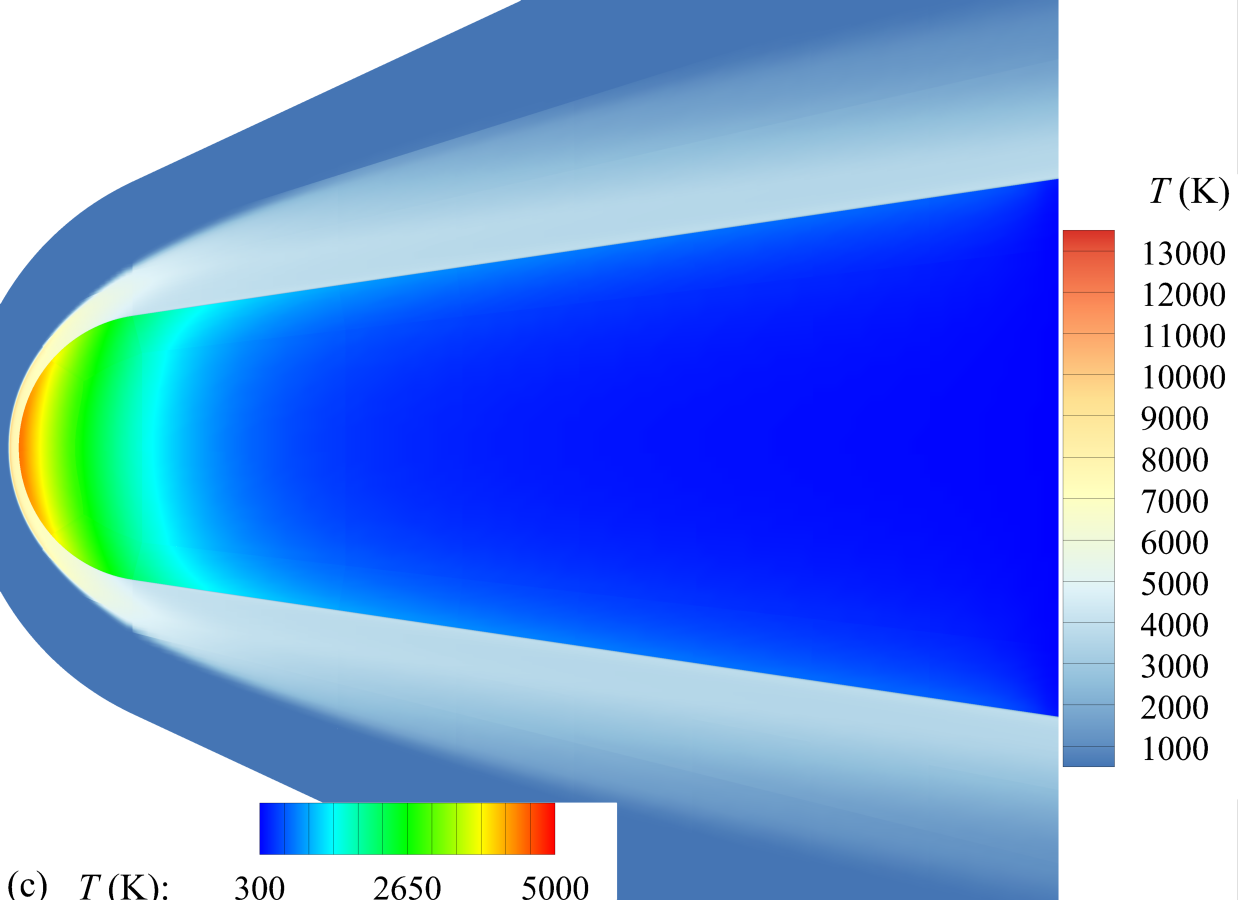}
	}	
	\subfigure
	{
		\includegraphics[width=0.45\textwidth]{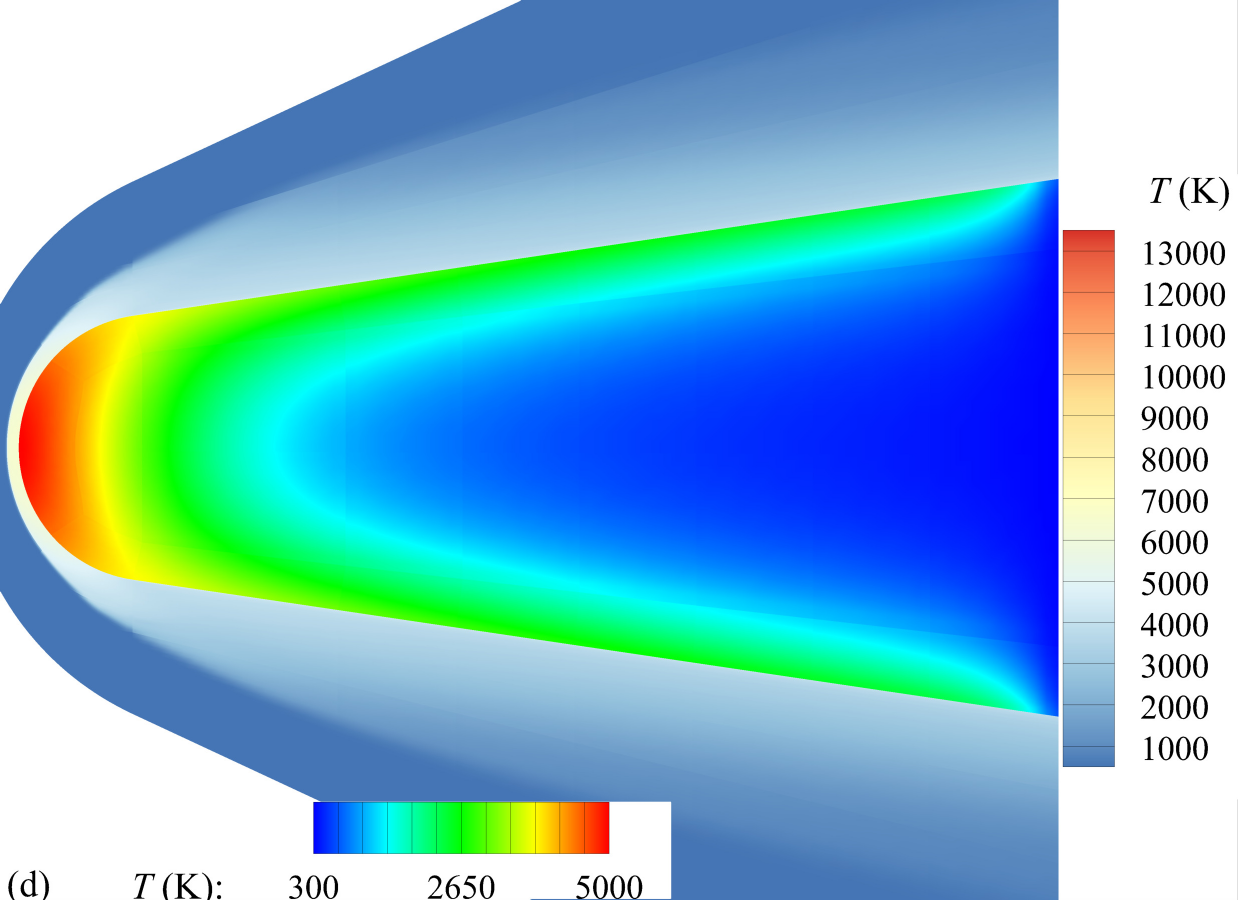}
	}
		% \centering
		% \includegraphics[width=0.6\textwidth]{figure/coupling.pdf}
	\caption{The temperature contour on the symmetry plane for the coupled CRF and MTR simulation of the IRV-2 reentry at different moment:(a)5s;(b)10s;(c)15s;(d)20s.}
	\label{Fig.IRV2-T}
\end{figure}

\begin{table}[htb!]
	\centering
	\caption{Free-stream condition for the reentry trajectory of the IRV-2 vehicle.}
	\label{Tab.trajectory}%添加标题 设置标签
	\begin{tabular}{cccc}
		\toprule
		%\hline
		Trajectory point&Time, s&Altitude, km&Mach number	\\
		1&0.0  	&66.9	&22.41	\\
		2&4.25 	&55.8	&21.08	\\
		3&6.75 	&49.2	&20.57	\\
		4&8.75 	&44.0	&20.90	\\
		5&10.25	&40.1	&21.29	\\
		6&11.50	&36.8	&21.58	\\
		7&12.50 &34.2	&21.78	\\
		8&13.25	&32.2	&21.91	\\
		9&13.95	&30.5	&21.84	\\
		10&14.75	&28.2	&21.7 	\\
		11&15.50	&25.7	&21.50	\\
		12&16.25	&22.8	&21.17	\\
		13&17.00	&19.7	&20.64	\\
		14&17.75	&16.3	&19.77	\\
		15&18.25	&13.9	&19.08	\\
		16&18.50	&12.7	&18.70	\\
		17&18.75	&11.5	&18.30	\\
		18&19.00	&10.3	&17.69	\\
		19&19.50	&7.89	&16.25	\\
		20&20.00	&5.53	&14.88	\\
		21&20.50	&3.27	&13.58	\\
		22&21.00	&1.13	&12.38	\\
		23&21.28	&0.00	&11.75	\\
		%\hline
		\bottomrule
	\end{tabular}
\end{table}
%In the coupled CRF and MTR simulation of the IRV-2 reentry, the physical time step is 0.25s, and the free stream conditions at different moment are taken from flight conditions of the trajectory. The CRF and MTR are coupled by the GSI interface through the explicit and implicit coupling mechanisms.
The wall temperature and the heat flux over time at different positions are shown in Fig.~\ref{Fig.IRV2coupling}. The wall heat flux of the explicit coupling mechanism deviates from the result of the implicit coupling method at 18 s. Thus, the wall temperature on the nose calculated by the implicit coupling solution reaches its maximum value at about 18s, while the wall temperature on the nose calculated by the explicit coupling solution reaches its maximum value at about 20s. It can be concluded that the implicit coupling mechanism decreases time discretization error and yields more reliable results. The temperature contour of the IRV-2 on the symmetry plane of the flow fields and the TPS material is shown in Fig.~\ref{Fig.IRV2-T}, the highest temperature in the flow fields is about 13000 K after the shock, while the highest temperature of the solid material is about 5500 K on the nose. The temperature in the flow fields are different over time due to the flight conditions and the wall temperature are changed during the flight trajectory. The flight velocity decreases rapidly at the end of the trajectory. Thus, the wall heat flux and the wall temperature of the IRV-2 vehicle are decreased after 18 s.

% \subsection{Reentry capsule}
% Finally, to investigate the ablation behavior of the TPS material during the Mars atmospheric entry, this study simulates the material thermal response of the Mars Science Laboratory (MSL) spacecraft in the entry aerothermal environment with the presence of heterogeneous surface reactions and pyrolysis. In addition, research has been conducted on the ablation flow field and thermal loads on the surface of the MSL. The MSL spacecraft is a 4.5 m diameter blunted aeroshell, which utilizes a ablative material with a total thickness of 31.75 mm, known as the PICA.  

\section{Conclusion}
In this paper, we propose a multicomponent material thermal response (MTR) solver to clarify the heat and mass transfer characteristics in the charring ablative materials. The governing equations of the MTR are based on the volume averaged multiphase conservation equations for mass, momentum, and energy. The fidelity of the numerical methods is enhanced by considering the complex composition of the pyrolysis gases, which increases the accuracy of the calculation of transport coefficients and thermodynamic parameters. To improved the computational efficiency, implicit time integration for the MTR solver is derived. The solution variables of the MTR solver can be solved independently via an explicit coupling mechanism, improving efficiency without sacrificing accuracy. The reliability of the MTR solver is verified by serval standard validation cases from TACOT. The multicomponent model computes higher temperature than the mix gas model in the backup region of the charring materials.

A gas-surface interaction (GSI) interface is proposed to couple the MTR solver with the chemically reacting flow solver to simulate the unsteady heating process of the TPS materials on hypersonic vehicles during its flight trajectory. The GSI interface incorporates surface mass balance equation and surface energy equation to account for the mass injection and energy dissipation on the ablating surface. Explicit and implicit coupling mechanism are developed depending on whether the GSI interface is updated at the physical iteration or the pseudo iteration. The explicit coupling mechanism updates interfacial quantities at physical time step achieving high efficiency but shows oscillations and time discretization errors. The implicit coupling mechanism addresses this issue by communicating interfacial quantities before each pseudo time step, thereby eliminating the time discretization errors and numerical oscillations. Consequently, the implicit coupling mechanism produces more reliable solutions of the coupled simulation between chemically reacting flows and material thermal response.
%\section*{Appendix}
%An Appendix, if needed, should appear before the acknowledgments.
\section*{Acknowledgments}

This work was supported by 111 project on ``Aircraft Complex Flows and the Control'' [grant number B17037].

\section*{Appendix A. Supplementation of the implicit scheme of the MTR}

The coordinate transformation mapping from physical space $(x, y, z)$ to the generalized space $(\xi,\eta,\zeta)$ is represented as 
\begin{equation}
	\left\{ {\begin{array}{*{20}{c}}
		{\xi  = \xi (x,y,z)}\\
		{\eta  = \eta (x,y,z)}\\
		{\zeta  = \zeta (x,y,z)}
		\end{array},J = \left| {\begin{array}{*{20}{c}}
		\xi_x&\xi_y&\xi_z\\
		\eta_x&\eta_y&\eta_z\\
		\zeta_x&\zeta_y&\zeta_z
		\end{array}} \right|} \right.
\end{equation}

The governing equations of the MTR in the $(\xi,\eta,\zeta)$ coordinates are given by,
\begin{equation}
	\frac{{\partial {\bm{\tilde{W}}}}}{{\partial t}} + \bm{R}(\bm{W}) = 0, \bm{R}(\bm{W})  = \frac{{\partial {\bm{\tilde{F}}}}}{{\partial \xi}} +\frac{{\partial {\bm{\tilde{G}}}}}{{\partial \eta}} + \frac{{\partial {\bm{\tilde{H}}}}}{{\partial \zeta}} + {\bm{\tilde{S}}},
\end{equation}
where the $\bm{\tilde{W}}, \bm{\tilde{S}}$ are the vector of the solution variables and the source term, respectively. $\bm{\tilde{F}}, \bm{\tilde{G}}, \bm{\tilde{H}}$ are the vector of the fluxes in the $(\xi,\eta,\zeta)$ direction. These formula are given by,

\begin{equation}
	\bm{\tilde{W}} = \frac{1}{J}\left[
	\begin{array}{c}
		p_1 \\
		\ldots \\ 
		p_{ns} \\ 
		\kappa T
	\end{array}\right],
	\bm{\tilde{F}} = \frac{1}{J}\left[
	\begin{array}{c}
		(\xi_xp_x+\xi_yp_y+\xi_zp_z)K_1  - pD_{1}\frac{\partial Y_1}{\partial \xi} \\ 
		\ldots \\
		(\xi_xp_x+\xi_yp_y+\xi_zp_z)K_{ns}- pD_{ns}\frac{\partial Y_{ns}}{\partial \xi}  \\ 
		\rho_{pg}Uc_{p,pg}T + q_{x}\xi_x+q_{y}\xi_y+q_{z}\xi_z
	\end{array}\right]
\end{equation}
\begin{equation}
	\bm{\tilde{G}} = \frac{1}{J}\left[
		\begin{array}{c}
			(\eta_xp_x+\eta_yp_y+\eta_zp_z)K_1 - pD_{1}\frac{\partial Y_{1}}{\partial \eta}\\ 
			\ldots \\
			(\eta_xp_x+\eta_yp_y+\eta_zp_z)K_{ns} - pD_{ns}\frac{\partial Y_{ns}}{\partial \eta} \\ 
			\rho_{pg}Vc_{p,pg}T + q_{x}\eta_x+q_{y}\eta_y+q_{z}\eta_z
		\end{array}\right]
	\end{equation}
	\begin{equation}
		\bm{\tilde{H}} = \frac{1}{J}\left[
			\begin{array}{c}
				(\zeta_xp_x+\zeta_yp_y+\zeta_zp_z)K_1 - pD_{1}\frac{\partial Y_{1}}{\partial \zeta} \\ 
				\ldots \\
				(\zeta_xp_x+\zeta_yp_y+\zeta_zp_z)K_{ns}- pD_{ns}\frac{\partial Y_{ns}}{\partial \zeta} \\ 
				\rho_{pg}Wc_{p,pg}T + q_{x}\zeta_x+q_{y}\zeta_y+q_{z}\zeta_z
			\end{array}\right],
		\bm{\tilde{S}} = \frac{1}{J}\left[
			\begin{array}{c}
				\frac{R_uT}{\epsilon_{pg}M}\frac{\partial \rho_1}{\partial t} \\ 
				\ldots \\ 
				\frac{R_uT}{\epsilon_{pg}M}\frac{\partial \rho_{ns}}{\partial t} \\ 
				\frac{\partial (\epsilon\rho e)_{pm}}{\partial t} + \frac{\partial (\epsilon\rho h)_{pg}}{\partial t}+ \epsilon_{pg}\frac{\partial p}{\partial t}
			\end{array}
		\right]
\end{equation}
where $\kappa = \rho_{pm}\epsilon_{pm}c_{v,pm}+ \rho_{pg}\epsilon_{pg}c_{v,pg}, K_1=\frac{p_1 K}{\epsilon_{pg} \mu}, K_{ns}=\frac{p_{ns} K}{\epsilon_{pg} \mu}, U = u\xi_x + v\xi_y + w\xi_z +\xi_t, V = u\eta_x + v\eta_y + w\eta_z +\eta_t, W = u\zeta_x + v\zeta_y + w\zeta_z +\zeta_t,p_x = \frac{\partial p}{\partial \xi}\xi_x + \frac{\partial p}{\partial \eta}\eta_x+\frac{\partial p}{\partial \zeta}\zeta_x,p_y = \frac{\partial p}{\partial \xi}\xi_y + \frac{\partial p}{\partial \eta}\eta_y+\frac{\partial p}{\partial \zeta}\zeta_y,p_z = \frac{\partial p}{\partial \xi}\xi_z + \frac{\partial p}{\partial \eta}\eta_z+\frac{\partial p}{\partial \zeta}\zeta_z, q_x = -\lambda\frac{\partial T}{\partial x}, q_y = -\lambda\frac{\partial T}{\partial y},q_z = -\lambda\frac{\partial T}{\partial z}$.

%The numerical framework employs finite difference method. Spatial discretization is implemented by a 2rd central difference scheme in generalized space. To improve numerical stabilities, the implicit time integration is achieved by the linearized equations, and 
The flux jacobian in the $\xi$ direction is given by,
\begin{equation}
	\begin{array}{c}
		\frac{\partial \bm{R}}{\partial p_{s,i+1}} = -\frac{K}{J}[(\frac{\xi_xp_x + \xi_yp_y +\xi_zp_z}{2\epsilon_{pg} \mu})_{i}+p_{s,i}(\frac{\xi_x^2 + \xi_y^2 + \xi_z^2}{\epsilon_{pg} \mu})_{i+1/2}], \\
		\frac{\partial \bm{R}}{\partial p_{s,i}} = -\frac{K}{J}[-p_{s,i}(\frac{\xi_x^2 + \xi_y^2 + \xi_z^2}{\epsilon_{pg} \mu})_{i+1/2}-p_{s,i}(\frac{\xi_x^2 + \xi_y^2 + \xi_z^2}{\epsilon_{pg} \mu})_{i-1/2}], \\
		\frac{\partial \bm{R}}{\partial p_{s,i-1}} = -\frac{K}{J}[-(\frac{\xi_xp_x + \xi_yp_y + \xi_zp_z}{2\epsilon_{pg} \mu})_{i} +p_{s,i}(\frac{\xi_x^2 + \xi_y^2 + \xi_z^2}{\epsilon_{pg} \mu})_{i-1/2}], \\
		\end{array}
\end{equation}
where $(b_x, b_y, b_z)$ are the gradient of $p$.
% \begin{equation}
% 	b_x = \frac{\partial p}{\partial x} = \xi_x\frac{\partial p}{\partial \xi} + \eta_x\frac{\partial p}{\partial \eta} + \zeta_x\frac{\partial p}{\partial \zeta}.
% \end{equation}

For the energy equation, the flux jacobian of is represented as
\begin{equation}
	\begin{array}{c}
		\kappa J\frac{\partial \bm{R}}{\partial T_{i-1}}=(\rho_gc_{p,pg}U)_{i-1/2}-[\lambda(\nabla \xi)^2]_{i-1/2}\\ +  
		\frac{1}{4}[(\lambda(\eta_x \xi_x+\eta_y\xi_y+\eta_z \xi_z))_{j+1/2}-(\lambda(\eta_x \xi_x+\eta_y\xi_y+\eta_z \xi_z))_{j-1/2}]\\+\frac{1}{4}[(\lambda(\zeta_x \xi_x+\zeta_y\xi_y+\zeta_z \xi_z))_{k+1/2}-(\lambda(\zeta_x \xi_x+\zeta_y\xi_y+\zeta_z \xi_z))_{k-1/2}]\\
		\kappa J\frac{\partial \bm{R}}{\partial T_{i+1}}=(\rho_gc_{p,pg}U)_{i+1/2}-[\lambda(\nabla \xi)^2]_{i+1/2}\\ -
		\frac{1}{4}[(\lambda(\eta_x \xi_x+\eta_y\xi_y+\eta_z \xi_z))_{j+1/2}-(\lambda(\eta_x \xi_x+\eta_y\xi_y+\eta_z \xi_z))_{j-1/2}]\\-\frac{1}{4}[(\lambda(\zeta_x \xi_x+\zeta_y\xi_y+\zeta_z \xi_z))_{k+1/2}-(\lambda(\zeta_x \xi_x+\zeta_y\xi_y+\zeta_z \xi_z))_{k-1/2}] \\
		\kappa J\frac{\partial \bm{R}}{\partial T_{i}}=-(\rho_gc_{p,pg}U)_{i+1/2}-(\rho_gC_{p,g}U)_{i-1/2}+\lambda(\nabla \xi)^2_{i+1/2}+ \lambda(\nabla \xi)^2_{i-1/2}\\
	\end{array}
\end{equation}
where $\kappa = (\rho \epsilon c_v)_{pm} +  (\rho \epsilon c_p)_{pg}, \nabla \xi = \sqrt{\xi_x^2 + \xi_y^2 + \xi_z^2}$.

The flux jacobian of the source term is represent as 
\begin{equation}
	\begin{split}
		J\frac{\partial \bm{R}}{\partial p_{s}} = \sum_{i}^{N_p} \sum_{j}^{P_i} \theta_{i,j,k} \beta_{i,v}\rho_{i,v}F_{i,j}(1-\chi_{i,j})^{m_{i,j}}T^{n_{i,j}} A_{i,j} \exp{(-\frac{E_{i,j}}{RT})}\frac{E_{i,j}}{R T^2}\frac{M}{R \rho_{pg}} \\ 
	\kappa J\frac{\partial \bm{R}}{\partial T} = \sum_{i}^{N_p} \sum_{j}^{P_i} \theta_{i,j,k} \beta_{i,v}\rho_{i,v}F_{i,j}(1-\chi_{i,j})^{m_{i,j}}T^{n_{i,j}} A_{i,j} \exp{(-\frac{E_{i,j}}{RT})}\frac{E_{i,j}}{R T^2}e_{pm}
	\end{split}
\end{equation}

\bibliographystyle{elsarticle-num}
%\biboptions{longnamesfirst,angle,semicolon,square,numbers,sort&compress}

\bibliography{manu}


%\begin{thebibliography}{00}


%\end{thebibliography}
\end{document}